\documentclass[%
 aip,
 amsmath,amssymb,
 reprint,%
]{revtex4-1}
\usepackage{amsmath,amssymb}
\usepackage{dsfont}
\usepackage{graphicx}
\usepackage{hyperref}
\usepackage{physics}
\usepackage{amsmath,amssymb}
\usepackage{dsfont}
\usepackage{graphicx}
\usepackage{hyperref}
\usepackage{physics}
\usepackage{mathtools}
\usepackage{cancel}
\usepackage{bbold}
\usepackage{bm}
\usepackage{color} 
\usepackage[dvipsnames]{xcolor}
\usepackage[T1]{fontenc}
\usepackage{tcolorbox}

\usepackage{soul}

\newcommand{\ommin}{f_{\mathrm{min}}}
\newcommand{\ommax}{f_{\mathrm{max}}}
\newcommand{\fmin}{f_{\mathrm{min}}}
\newcommand{\fmax}{f_{\mathrm{max}}}

\definecolor{blue}{rgb}{0.0, 0.0, 1}

\definecolor{asparagus}{rgb}{0.53, 0.66, 0.42}

\begin{document}
\newcounter{theo}
\author{M. K\k{e}pa}
\affiliation{%
Institute of Physics, Polish Academy of Sciences, al.~Lotnik{\'o}w 32/46, PL 02-668 Warsaw, Poland
}%
\author{N. Focke}
\affiliation{%
JARA-FIT Institute for Quantum Information, Forschungszentrum J\"ulich GmbH and RWTH Aachen University, Aachen, Germany
}%
\author{{\L}. Cywi\'nski }
\affiliation{%
Institute of Physics, Polish Academy of Sciences, al.~Lotnik{\'o}w 32/46, PL 02-668 Warsaw, Poland
}%
\author{J. A. Krzywda}
 \email{krzywda@ifpan.edu.pl}
\affiliation{%
Institute of Physics, Polish Academy of Sciences, al.~Lotnik{\'o}w 32/46, PL 02-668 Warsaw, Poland
}
\affiliation{%
Lorentz Institute and Leiden Institute of Advanced Computer Science,
Leiden University, P.O. Box 9506, 2300 RA Leiden, The Netherlands
}

\title{Simulation of $1/f$ charge noise affecting a quantum dot in a Si/SiGe structure}
\begin{abstract}

Due to presence of magnetic field gradient needed for coherent spin control, dephasing of single-electron spin qubits in silicon quantum dots is often dominated by $1/f$ charge noise. 
We investigate theoretically fluctuations of ground state energy of an electron in gated quantum dot in realistic Si/SiGe structure. We assume that the charge noise is caused by motion of charges trapped at the semiconductor-oxide interface. We consider a realistic range of trapped charge densities, $\rho \! \sim \! 10^{10}$ cm$^{-2}$, and typical lenghtscales of isotropically distributed displacements of these charges, $\delta r \! \leq \! 1$ nm, and identify pairs $(\rho,\delta r)$ for which the amplitude and shape of the noise spectrum is in good agreement with spectra reconstructed in recent experiments on similar structures.
\end{abstract}
\maketitle

Charge noise with $1/f$ spectrum is ubiquitous in semiconductor nanostructures. \cite{duttaLowfrequencyFluctuationsSolids1981,Paladino_RMP14,fleetwoodNoiseDefectsMicroelectronic2015,islamBenchmarkingNoiseDephasing2022}  In gate-defined quantum dots (QDs) \cite{Wuetz_NC23} this noise is probably caused by bistable sources of electric fields localized at the semiconductor/insulator interface, or within the insulator separating semiconducting material from the gate.\cite{sakamoto_nakamura_nakamura_1995, liefrink_dijkhuis_houten_1994,ramonDecoherenceSpinQubits2010,bermeisterChargeNoiseSpinorbit2014} Spins of electrons or holes confined in QDs have been demonstrated to be viable qubits,\cite{Hanson_RMP07,Burkard21} with the gated QD architectures showing constant progress towards scalable multi-qubit devices.\cite{philipsUniversalControlSixqubit2022, Borsoi22} While using the spin of a single electron as the qubit makes its quantum state insensitive, to the first order, to charge noise, decoherence caused by that noise is relevant for operation of two-qubit gates \cite{huangSpinDecoherenceTwoqubit2018,  veldhorstTwoqubitLogicGate2015a, ZajacScience18, watsonProgrammableTwoqubitQuantum2018b} and evolution of singlet-triplet qubits based on two electrons in two QDs. \cite{Coish_PRB05,Hu_PRL06,Dial_PRL13,TakedaPRL2020,Connors_NC22,Chan_PRAP18,Qi_PRB17,culcerDephasingSiSinglettriplet2013} Moreover, charge noise has  been recently recognized as the factor limiting the quality of single-spin qubits in silicon-based QDs. \cite{Yoneda_NN18,Struck_NPJQI20,Saraiva_AFM21} This is due to presence of either real,
\cite{gilbert_tanttu_lim_feng_huang_cifuentes_serrano_mai_leon_escott_etal_2023,harvey-collard_jacobson_bureau-oxton_jock_srinivasa_mounce_ward_anderson_manginell_wendt_etal_2019, Tanttu_PRX19}
or artificial spin-orbit couplings caused by presence of magnetic field gradients, which are necessary for single-qubit operations.\cite{takeda_nat22,Takeda_SA16,Brunner_PRL11,ZajacScience18,Yoneda_NN18}

The microscopic sources of charge noise in gated QD architectures have not been fully characterized yet. However, recent progress in reconstruction 
of spectra of $1/f$ charge noise affecting the QDs \cite{Struck_NPJQI20,Connors_PRB19,Connors_NC22,Yoneda22} 
has provided a lot of experimental data against which one could test various models of these sources. 
It is now known that the power spectral density (PSD) of noise of electronic energy levels in a typical QD has $1/f$ form for $f \in [10^{-3},10^6]$ Hz (and possibly in an even broader range), albeit with visible deviations from ideal $1/f$ scaling, and the typical amplitude of noise at 1 Hz is of the order of $1$ $\mu$eV/$\sqrt{\mathrm{Hz}}$ for electrons in Si/SiGe,\cite{Connors_NC22,Struck_NPJQI20,Wuetz_NC23,petersson_petta_lu_gossard_2010,Mi_PRB18} while being within one order of magnitude from this value for any other characterized spin qubit, see Ref.~\onlinecite{Wuetz_NC23} and references therein.

In recent papers modeling of charge noise was performed for electron\cite{Shehata22} and hole\cite{Shalak22} spin qubits in SiMOS devices. Here we focus on electron spin qubits in Si/SiGe devices of the type used in recent experiments on electron spin qubit shuttling.\cite{Seidler_NPJQI22,Langrock22} 
We assume that the noise is caused by sub-nm motion of charges trapped at the Si/SiO$_{2}$ interface \cite{klosCalculationTunnelCouplings2018, klosSpinQubitsConfined2019, spruijtenburgPassivationCharacterizationCharge2016, thoanInterfaceStateEnergy2011,Campbell05,
Langrock22} that have density $\rho_{\mathrm{ic}} \! \sim \! 10^{10}$ cm$^{-2}$. Simulations in Ref.~\onlinecite{Langrock22} showed that using $\rho_{\mathrm{ic}} \! =\! 5 \times 10^{10}$ cm$^{-2}$ reproduces typically observed electrostatic-disorder-induced energy detunings between neighboring QDs. Focusing on $\rho_{\mathrm{ic}}$ in range from $5\times 10^{9}$cm$^{-2}$ to $10^{11}$cm$^{-2}$, we calculate the influence of these charges on the ground-state (GS) orbital energy of an electron in a  QD in the considered nanostructure. We then investigate what typical length scale of isotropic charge motion and charge densities lead to reproduction of measured shape and amplitude of $1/f$-like noise spectra. 

\begin{figure}[tb!]
    \centering
    \includegraphics[width=1\columnwidth]{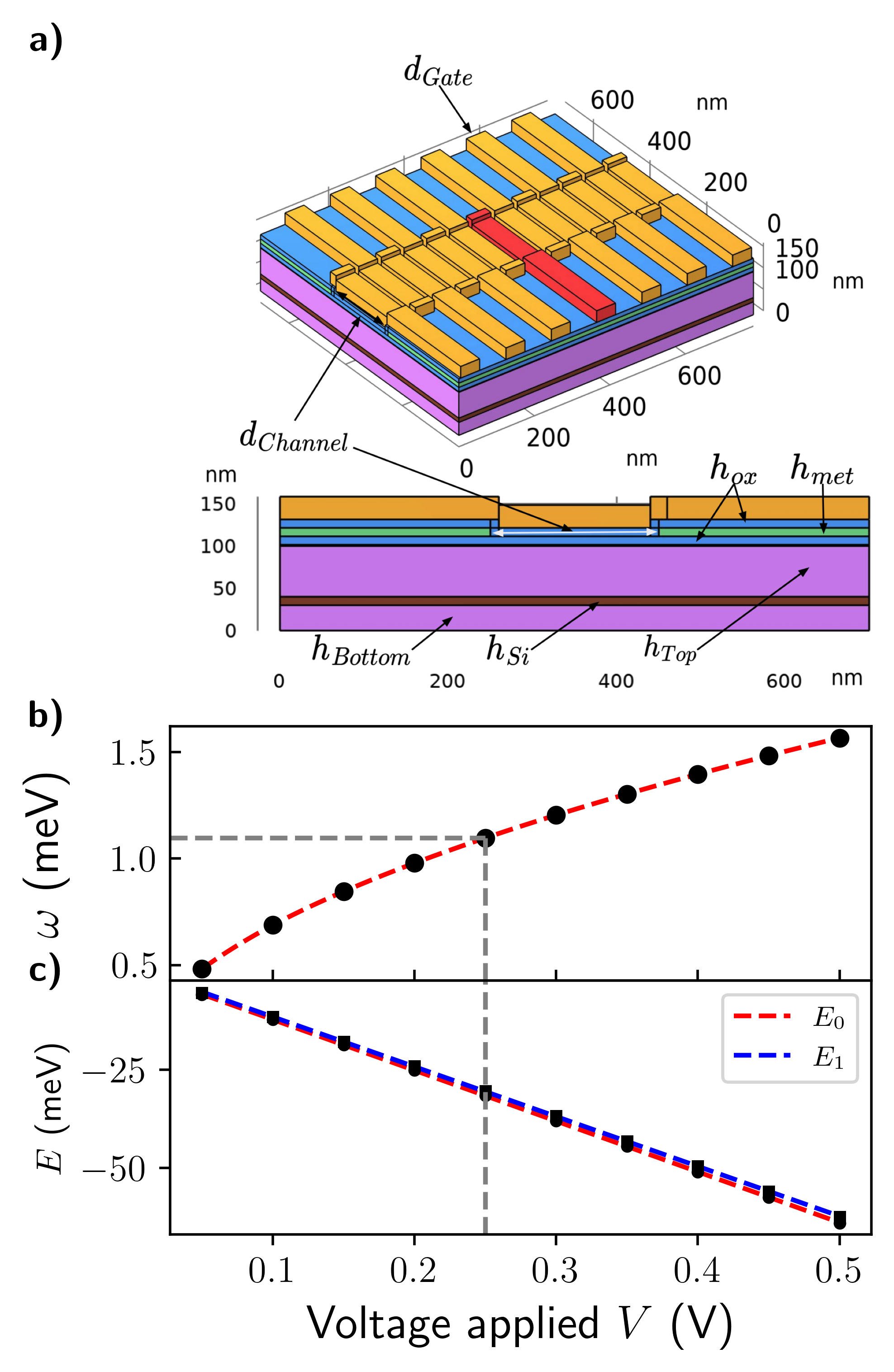}
    \caption{a) CAD model of the device. Electrodes that define the potential located on top of the device are separated from the bulk structure with conductive and insulating oxide layers. $d_{Gate} = 50 \ \text{nm}$, $d_{Channel} = 60 \ \text{nm}$, $h_{Top} = 60 \ \text{nm}$, $h_{Si} = 10 \ \text{nm}$, $h_{Bottom} = 30 \ \text{nm}$. b) Orbital gap $\omega = E_1 - E_2$ as a function of applied voltage with interpolated trend line. Grey dashed line marks $V=0.25\ \text{V}$ that corresponds to $\omega \approx 1.1\,
\text{meV}$, which is used in all further simulations. c) The dependence of individual energy levels along with lever arm fits with $\alpha_0 = -128.86\ \text{meV/V}$ and $\alpha_1 = -126.54\ \text{meV/V}$.} 
    \label{fig:fig1}
\end{figure}

We employ the simplest model leading to $1/f$ noise spectrum: that of many {\it independent} sources of Random Telegraph Noise (RTN) \cite{Machlup_JAP54, ramonDecoherenceSpinQubits2010, Constantin_PRB09} 
- the so-called two-level fluctuators (TLFs)- 
having switching rates $\gamma$ drawn from a log-normal distribution spanning over many orders of magnitude of $\gamma$. \cite{schriefl2006decoherence,YouPRR21, Shnirman_PRL05,Muller_2019} 
We consider here noise at fixed temperature, i.e.~$T\approx \! 100$ mK at which spin qubits are typically operated, and assume equal average occupations of each state of a TLF. This means without considering the origin of TLFs dynamics (see e.g.~\onlinecite{Beaudoin_PRB15,Ahn_PRB21} for discussion of some of the models).
Distinct electric fields generated by the TLF in each of its states cause state-dependent shifts of ground state energy of the nearby QD. We define the coupling of the QD to the $n$-th TLF, $\eta_n$, as the shift of the energy of the GS orbital of an electron in the QD caused by a switching of the TLF. 
We focus on switching events that correspond to a sub-nm motion of a defect charge in the insulating layer between the semiconductor and the metallic gate (the blue region in Fig.~\ref{fig:fig1}). The two locations of the charge corresponding to the two states of the $n$-th TLF are $\mathbf{r}_n$ and $\mathbf{r}_n+\delta \mathbf{r}_n$. We have checked that charge tunneling between the insulator and the metal leads to shifts of energy levels of the QD that are too large to be reconciled with the experimentally observed noise levels, unless defect is less than $0.2$ nm away from the metal.
As the tunneling events were also excluded as a source of charge noise in similar analysis performed for SiMOS devices,\cite{Shehata22}, here we only take into account the motion of defect localized at semiconductor-oxide interface.

The PSD of the noise in GS energy of the electron in the QD is given by a sum of Lorentzian PSDs of the RTNs:
\begin{equation}
\label{eq:s_tls}
    S(f) = \sum_{n=1}^{N} \frac{2\eta_n^2}{\gamma_n + (2\pi f)^2/\gamma_n} \,\, . 
\end{equation}
For a large number of TLFs, a flat distribution of $\eta_n$, and a log-normal distribution of $\gamma_n$, i.e.~$p(\gamma_n) \propto 1/\gamma_n$, results in the PSD that is well approximated by the 1/f noise spectrum, i.e. $S(f) \! \approx \! S_1 f_1 / f$,
where $S_1$ is the power of the noise at $f_1 = 1$Hz, which is of the order of $1$ $\mu$eV$^{2}$/Hz in state-of-the-art Si/SiGe QDs.\cite{KranzAM20} 

\begin{figure}[tb!]
    \centering
    \includegraphics[width=1\columnwidth]{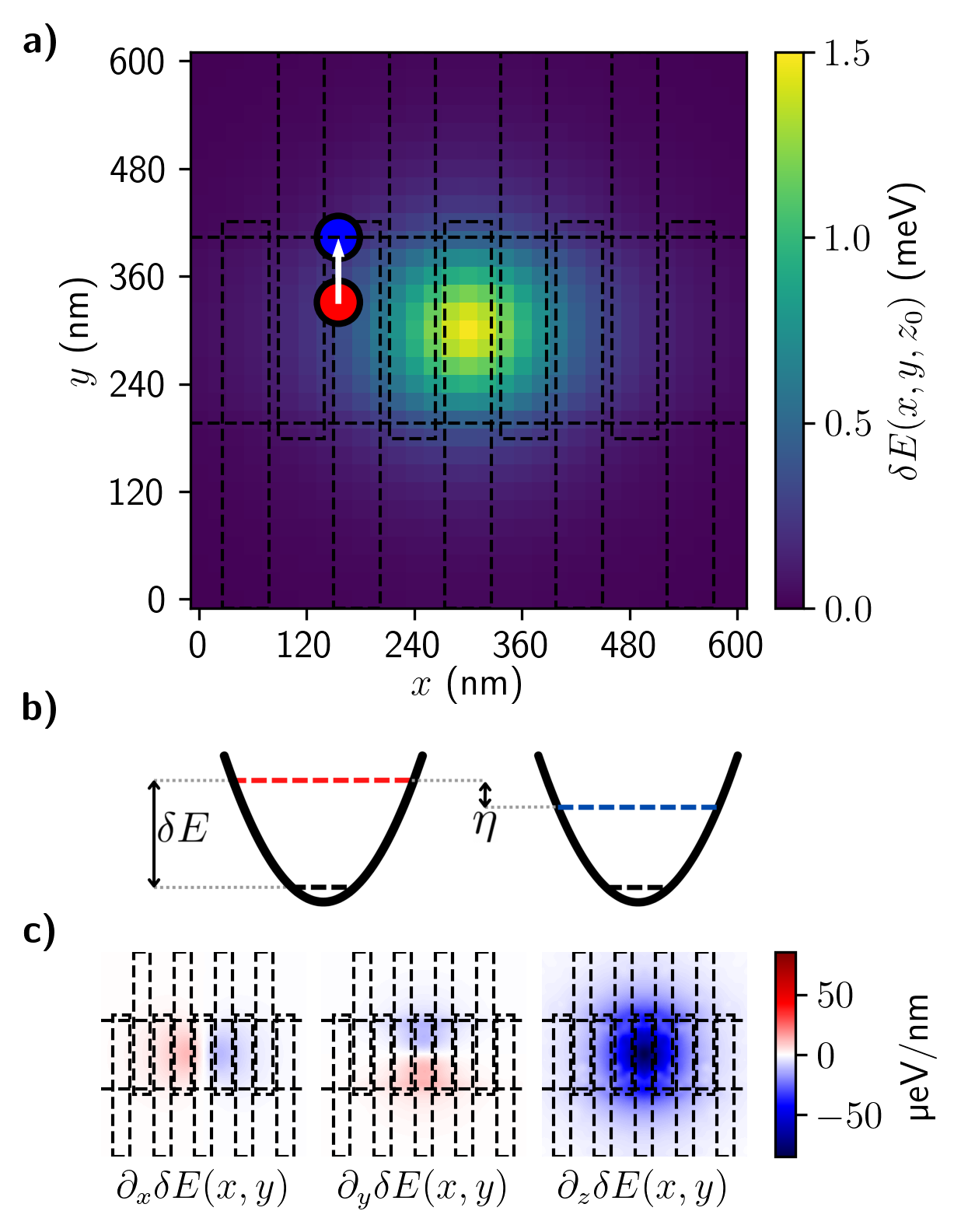}
 \caption{Results of the finite-element method implemented in COMSOL. a) Shift of ground state energy $\delta E(\mathrm{r})$ due to charge defect located at $\mathrm{r} = (x,y,z_0)$, where $z_0 = 102$nm (see Fig.~\ref{fig:fig1}a). b) Illustration of the energy shift $\delta E$ (left) caused by the presence of a point charge, and additional shift $\eta$ (right) caused by a displacement of this charge from red to blue point in panel a). We exaggerate the displacement size for illustration purposes. c) Interpolated spatial derivatives of $\delta E(\mathbf{r})$ along all three dimensions.}
    \label{fig:fig2}
\end{figure} 

We consider a realistic Si/SiGe device based on Refs.~\onlinecite{Langrock22,Seidler_NPJQI22}, the CAD model which is shown in Fig. \ref{fig:fig1}a. The structure consists of a quantum well (pink) that has two $\text{Si}_{0.7}\text{Ge}_{0.3}$ layers with a 10 nm thick Si layer in between. On top, there are 13 electrodes (yellow, red),to which voltages are applied. We use a gate pitch of $60\,\mathrm{nm}$ and the width of each gate is set to $50\,\mathrm{nm}$, which are within the range considered in Ref.~\onlinecite{Langrock22}. In this  design, we have a 10 nm screening layer (green) that is insulated from the quantum well and the electrodes by another 10 nm of the oxide layer (blue), followed by a 1.5 nm Si Cap layer. We use $\epsilon_r \! = \! 3.9$ for the relative permittivity of the SiO$_2$ oxide.
We define a channel as a gap in the screening layer, which allows the electric field to penetrate the structure and define the QDs about 80 nm below the terminal. We set $d_\text{Channel} = 200\,\mathrm{nm}$ as in Refs.~\onlinecite{Langrock22,Mills_SA22}. The remaining dimensions are given in the caption of Fig.~\ref{fig:fig1}a. The device structure and dimensions are similar to those of experimentally studied QDs,\cite{zajac_hazard_mi_wang_petta_2015,Connors_NC22,Borselli_2015,Wuetz_NC22,holman_rosenberg_yost_yoder_das_oliver_mcdermott_eriksson_2021}, although except for Ref.~\onlinecite{holman_rosenberg_yost_yoder_das_oliver_mcdermott_eriksson_2021} plunger and barrier gates tend to be located on separate layers. We neglect any influence of electron reservoirs, as due to possibility of charge shuttling \cite{Seidler_NPJQI22,Langrock22} they can be located sufficiently far from the QDs used as qubits.

We consider a single QD defined under the central gate (red).  We solve for electrostatic potential with voltages set on the electrodes, and we then find the electron GS energy, $E_{0}$, by solving the 3-dimensional Schroedinger equation for the envelope function in the 10 nm Si layer. We have used single-band effective mass approximation and omitted spin and valley degrees of freedom. We characterize the QD in Fig.~\ref{fig:fig1}bc, where the orbital gap and energies of the two lowest orbitals are shown as a function of the voltage applied to central gate, $V$, while keeping the other gates grounded. We choose $V=0.25$ V giving a realistic orbital gap $\omega = E_1 - E_0 \approx 1.1$ meV.

We assume that the charge noise is generated by the motion of negative charges close to the semiconductor/oxide interface about 60 nm above the Si quantum well, i.e.~on the upper surface of pink block at $z_0 \approx 100$ nm in Fig.~\ref{fig:fig1}a. A charge at $\mathbf{r}_n = (x_n,y_n,z_0)$ shifts the GS of the electron by $\delta E(\mathbf{r}_n) = E_0'-E_0$, where $E_0'$, $E_0$ are the GS energies with, and without the charge present, respectively. To compute this shift we insert a point charge $(-|e|)$ at $\mathbf{r}_n = (x_n,y_n,z_0)$, and repeat the procedure described above: compute the electrostatic potential, and find the new GS energy $E_0' = E_0+\delta E(\mathbf{r}_n)$. In Fig.~\ref{fig:fig2}a we plot $\delta E(\mathbf{r}_n)$ for the device shown in Fig.~\ref{fig:fig1}. We assume that the charges are located near the $z_0 = 102$ nm plane, i.e.~$0.5$ nm above the interface between the oxide and the Si cap layer. The results for $A = 600$ nm x $600$ nm region are obtained using COMSOL software, with a regular grid of 31x31 charge positions in the plane. We have checked that shifts from two charges localized at $\mathbf{r}_n$ and $\mathbf{r}_m$ are approximately additive, i.e. $\delta E(\mathbf{r}_n,\mathbf{r}_m) \approx \delta E(\mathbf{r}_n)+\delta E(\mathbf{r}_m)$ within a $2\%$ relative error. 

\begin{figure}[htb!]
\includegraphics[width=0.9\columnwidth]{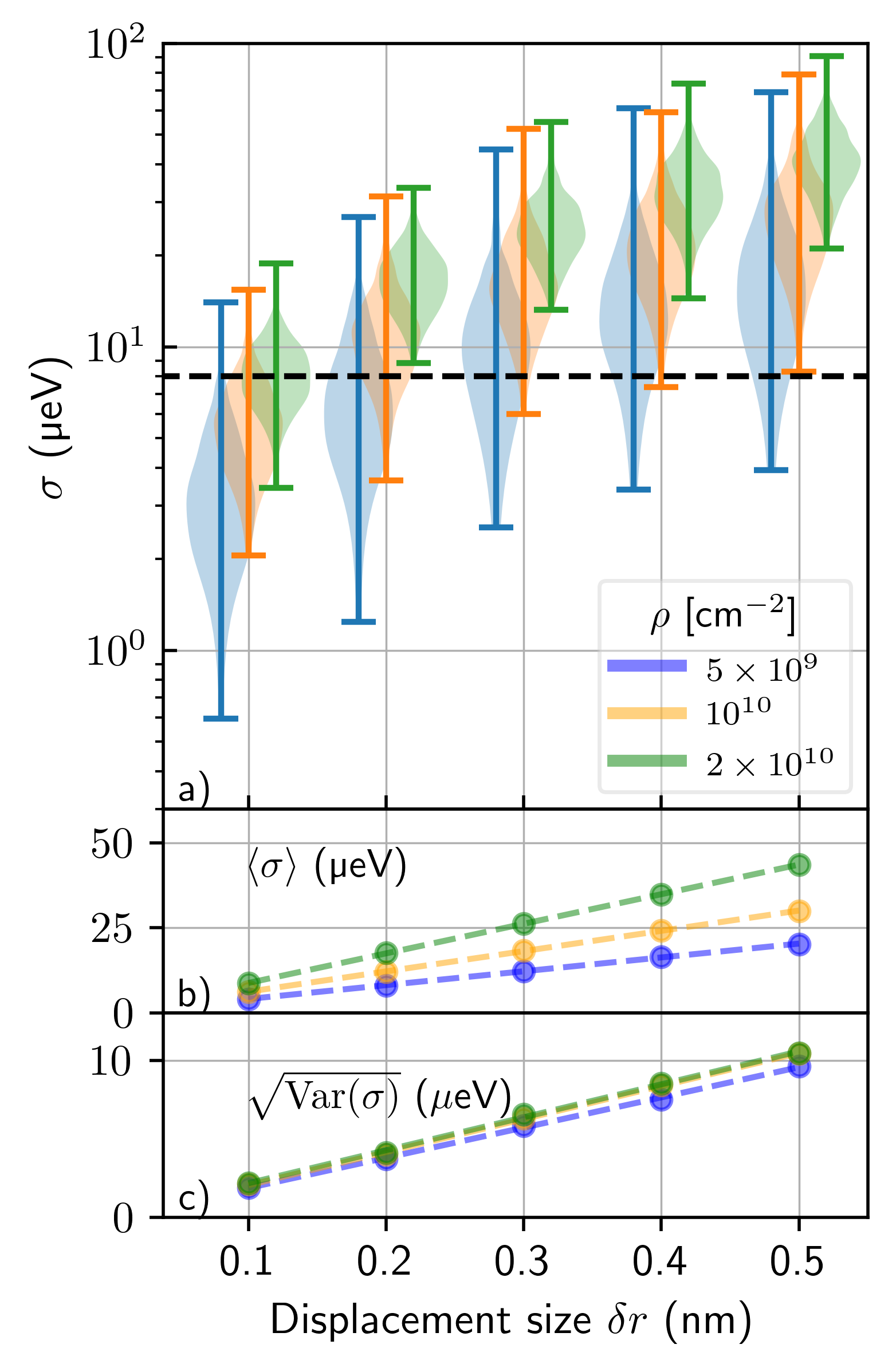}
\caption{Statistics of the amplitude of the noise, for different charge densities $\rho$ and magnitudes of charge displacement, $\delta r$. a) The distribution of noise amplitude from 1000 realizations for $\delta r = 0.1,0.2,0.3,0.4,0.5$nm (x-axis) and $\rho = 5\times 10^9,10^{10},2\times 10^{10}$cm$^{-2}$ (colors). Black dashed line corresponds to the targeted $\sigma_0 =  8\mu$eV. b) Ensemble averaged noise amplitude, $\langle \sigma \rangle$ and c) standard deviation of noise amplitude over the ensemble $\sqrt{\text{Var}(\sigma)}=  \sqrt{\langle \sigma^2 \rangle - \langle \sigma \rangle^2}$, both as a function of $\delta r$ for three charge densities from a), with linear trends marked by dashed lines).
}
\label{fig:fig3}
\end{figure}

To model a single TLF we add a random displacement of the charge position, $\delta \mathbf{r}_n$ that causes an additional shift in the GS energy (see Fig.~\ref{fig:fig2}), given by 
\begin{equation}
    \eta_n = \delta E(\mathbf{r}_n+\delta \mathbf{r}_n) - \delta E(\mathbf{r}_n) \,\, ,
\end{equation}
defining the coupling between the TLF and the QD. As the relative energy shifts due to $|\delta \mathbf{r}_n|\! < \! 1$ nm  are small, we approximate the coupling by
\begin{equation}
\label{eq:eta_def}
\eta_n \approx  \nabla  \delta E(\mathbf{r}_n) \cdot  \delta \mathbf r_n,
\end{equation}
where $\nabla = (\partial_x, \partial_y, \partial_z)$. In Fig.~\ref{fig:fig2}c we show the components of $\nabla\delta E(\mathbf{r})$, generated from interpolated $\delta E(\mathbf{r})$. For slowly varying $x$ and $y$ derivatives we interpolate results from Fig.~\ref{fig:fig2}a. For $z$ derivative we interpolate results obtained at $z_0 = 101.75,102.0,102.25,102.5$nm planes.

The motion of the charge away from the dot leads to decrease of $\delta E(\mathbf{r})$, i.e.~$\eta < 0$. In the $xy$ plane this is due to change of a distance between the defect and a QD, which leads to $|\partial_{x,y} \delta E(\mathbf{r})| \approx 10 \mu$eV/nm in the central region of $< \! 100$ nm from the dot. The motion towards and away from the dot results in  the opposite sign of $x$ and $y$ derivatives on the two sides of the QD center. 
On the other hand, the derivative in $z$ direction is always negative, and about an order of magnitude larger, as it is related to a decrease of the distance between the defect and metallic layer, i.e.~a decrease of a dipole moment of the electron and its mirror charge in the metallic gate. As a result, $\partial_z \delta E(\mathbf{r}) \approx 50\mu$eV/nm in the central region of $\approx 100$ nm radius, in which $\delta E(\mathbf{r})$ is significant. Outside of this region $\partial_z \delta E(\mathbf{r})$ decays towards $\approx 0.5\mu$eV/nm at the edge of the considered region $A$. This shows that the TLFs outside of $A$ do not contribute significantly to the energy shift.

 We compute the PSD of total noise  by adding contributions of all the TLFs, each given by Eq.~(\ref{eq:s_tls}).
 This approach is consistent with $\eta_n\ll \delta E(\mathbf{r}_n)$, and additivity of $\delta E(\{\mathbf{r}\})$, which has been  tested for two charges. We define $N$ as the number of TLFs that are active, i.e.~have their $\gamma_n$ in the range of frequencies $[\fmin,\fmax]$. We choose this range as $12$ order of magnitude in frequency, with $\fmin = 10^{-6}\,$Hz and $\fmax = 10^{6}\,$Hz. This safely covers the range in which $1/f$-like charge noise was observed in Si/SiGe QDs. \cite{Connors_PRB19,Struck_NPJQI20,Connors_NC22,Yoneda22} It is also roughly the range of frequencies that contribute to inhomogeneous broadening of splittings of QD-based qubits, for which $\fmin$ and $\fmax$ are given, roughly, by inverse of total data acquisition time, and single data point measurement time, respectively. 

 As a result the density of active TLFs, $\rho$, is upper-bounded by density of all the interface charges, $\rho_{\mathrm{ic}} \! \sim \! 10^{11}$ cm$^{-2}$    (see Ref.~\onlinecite{ZimmermannPRL1981}), as it is not known how many TLFs have $\gamma_n$ outside of the considered range. Note, however, that the log-normal distribution of $\gamma_n$ implies a constant number of TLFs per frequency decade, defined as $\text{fpd} = \ln(\fmax/\fmin)/N$. Finally, for a given $\rho$ we have $N\!= \!\rho A$, where $A \!= \! 600\text{nm}\times 600\text{nm}\approx  3.6\times 10^{-10} \text{cm}^2$  is the considered area near of the oxide-semiconductor interface, see Fig.~1.

The initial positions of the TLFs $\{\mathbf{r}_n\}$ are drawn from a uniform spatial distribution. As we assume the displacement vector of each RTN to be independent and isotropic, for each realization we draw components of $\delta \mathbf{r}_n$ from independent normal distributions,
    $(\delta \mathbf{r}_n)_i = \mathcal{N}(0,\delta r^2/3)$ for  $i\! = \! x,y,z$, 
so that the displacement size is $\delta r$.

To gauge our results against a measurable quantity, we first concentrate on the variance of the fluctuations of the GS energy, defined as:
\begin{equation}
\sigma^2 = 2\int_{\fmin}^{\fmax} \text{d}f S(f)  \approx 2\sum_n^{N} \eta_n^2.
\end{equation}
This quantity is related to the cohererence time of the qubit $T_{2}^{*} \! =\! \alpha \sqrt{2}/\sigma$, where for spin qubits $\alpha$ is the appropriate lever arm converting electric noise to spin splitting noise due to, e.g., presence of a magnetic field gradient.\cite{Yoneda_NN18,Struck_NPJQI20,Yoneda22} For charge qubits, $\alpha \! \approx \! 1$, and typically measured\cite{Shi_PRB13} $T_2^*\approx 0.1$ns gives $\sigma_0\approx 8$ $\mu$eV. This is confirmed by measured $1/f$ spectrum, with its value at $f_1=1$Hz given by $S_1 \approx 1\mu$eV$^2$/Hz \cite{KranzAM20}, as we have the relation
\begin{equation}
    \sigma^2 = 2\int_{\fmin}^{\fmax} \text{d}f S(f) = 2S_1 f_1 \ln(\tfrac{\ommax}{\ommin}) \approx (8\mu\text{eV})^2,
\end{equation}
where we have used $\fmax/\fmin = 10^{12}$. 

Let us now narrow down the range of possible combinations of $\rho$ and $\delta$r that give $\sigma_0 = 8$ $\mu$eV. In Fig.~\ref{fig:fig3}a we show numerically generated distributions of $\sigma$ as a function of $\delta r$ for different values of $\rho$. The considered range of $\delta r$ covers the values in the few $\AA$ range considered in Refs.~\onlinecite{Shehata22,culcerDephasingSiSinglettriplet2013, Reinisch06, biswasHydrogenFlipModel1999}.
The values of $\rho$ correspond to $N \approx 18, 36, 72$ charges in the considered region. In Fig.~\ref{fig:fig3}b we show that the typical noise amplitude, defined as the ensemble average of $\sigma$, scales linearly with $\delta r$, i.e. $\langle \sigma \rangle = a_\text{avg}(\rho)\delta r$. We highlight that larger $\delta r$ increases the ensemble variance of noise amplitude $\text{Var}(\sigma) = \langle \sigma^2\rangle - \langle\sigma \rangle^2$. From Fig.~\ref{fig:fig3}c we conclude that $\sqrt{\text{Var}(\sigma)}$ also scales linearly with $\delta r$, i.e. $\sqrt{\text{Var}(\sigma)} \approx a_\text{std}(\rho) \delta r$, however its dependence on $\rho$ is weak.

\begin{figure}[tb]
\includegraphics[width=1\columnwidth]{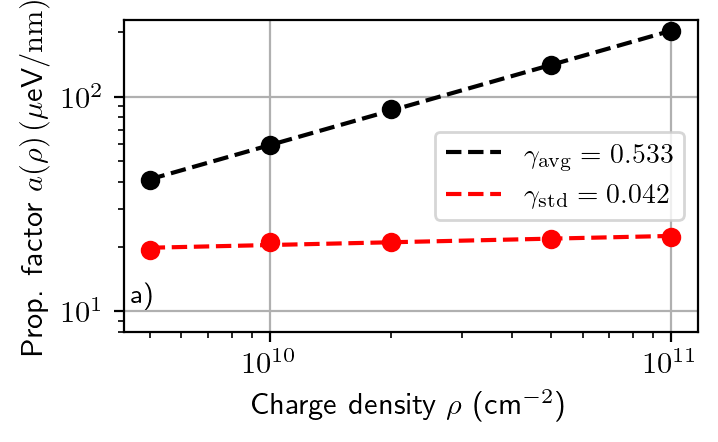}
\includegraphics[width=1\columnwidth]{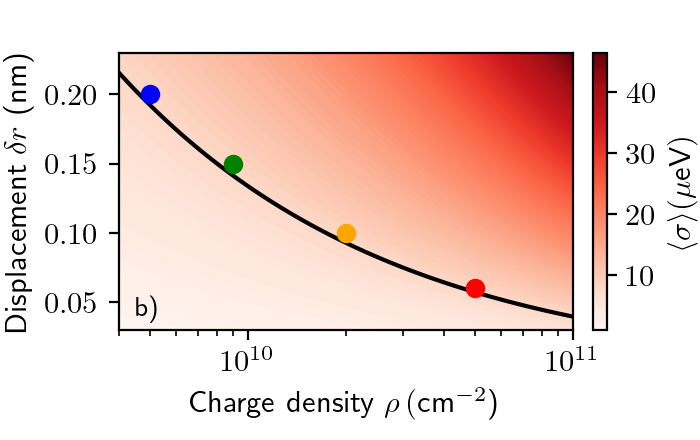}
\caption{Noise amplitude as a function of $\rho$ and $\delta r$. a) Proportionality factor of the ensemble average of noise amplitude $\langle \sigma \rangle \propto a_\text{avg}(\rho) \delta r$ (black) and corresponding ensemble variance $\sqrt{\langle \sigma^2 \rangle - \langle  \sigma\rangle^2} \propto a_\text{std}(\rho)\delta r$ (red). Points corresponds to the numerical results obtained by averaging 1000 realizations, dashed-lines are the linear fit on the log scale. From the fit we identify the corresponding exponents $a(\rho) \propto \rho^\gamma$ (see the legend for their values). b)  The fitted formula, Eq.~\eqref{eq:avg_sig}, as a function of $\rho$ and $\delta r$. Black line corresponds to $\langle \sigma \rangle \! = \! \sigma_0 \! = \! 8\mu$eV, and dots denote sets of parameters used to plot spectral densities in Fig.~\ref{fig:fig5}.}
\label{fig:fig4}
\end{figure}

\begin{figure}[tb]
\includegraphics[width=0.9\columnwidth]{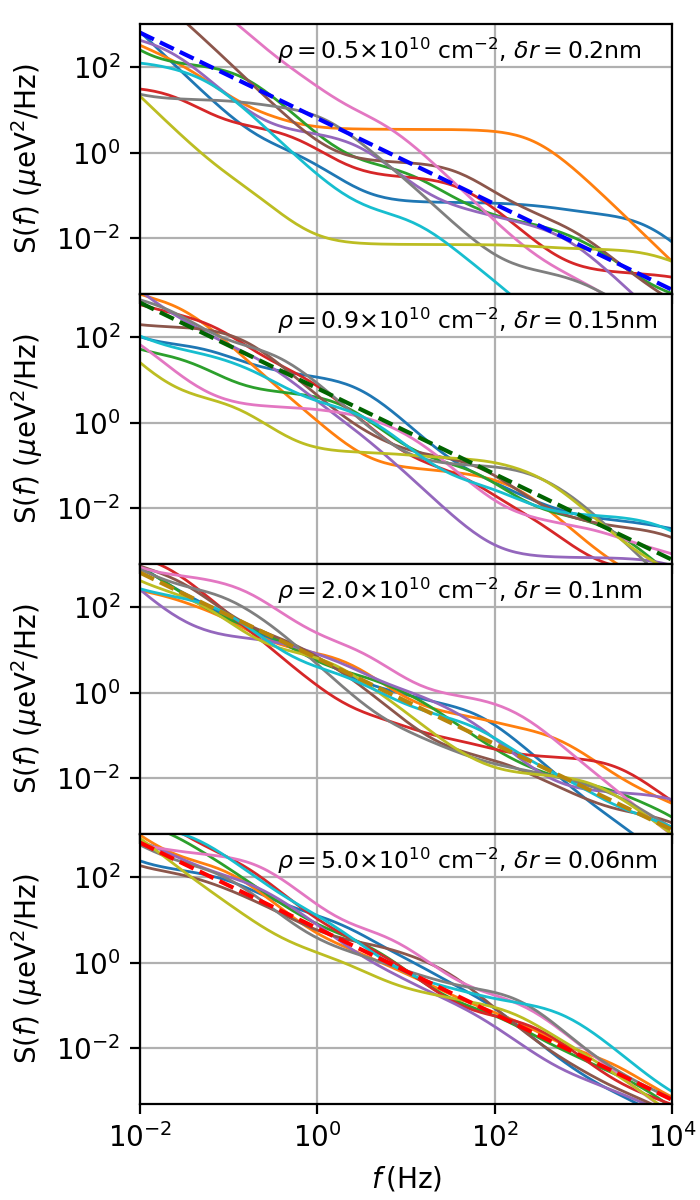}
\caption{Random realizations of power spectral densities for a selection of displacement magnitudes, $\delta r$, and charge densities, $\rho$, which result in a typical noise amplitude $\sigma_0$. The dashed lines are the ideal $1/f$ noise with $S_1 = 1\mu$eV$^2$/Hz, and their color is related to points marked in Fig.~\ref{fig:fig4} b. The solid lines are used to plot 10 realizations of TLF positions and their displacement for each set of parameters. }
\label{fig:fig5}
\end{figure}

We investigate the $\rho$-dependence of 
$a_\text{avg}(\rho)$ and $a_\text{std}(\rho)$ in Fig.~\ref{fig:fig4}a, where we plot them
for five values of $\rho$. The match between the results and the linear trend on the log-log scale allows us to extract the exponents $\gamma$ in $a(\rho) \propto \rho^\gamma$. Fitting gives an approximate expression for the ensemble-averaged amplitude:
\begin{equation}
\label{eq:avg_sig}
    \langle \sigma \rangle = (c_0\rho^{\gamma_\text{avg}} + a_{\text{avg},0})\delta r + b_0,
\end{equation}
where $c_0 = 2.77 \times 10^{-4} \tfrac{\text{cm}}{\text{nm}} \mu$eV, $\gamma_\text{avg}=0.533$, $ a_{\text{avg},0} =  0.951 \mu$eV/nm and $b_0$ is negligibly small. In Fig.~\ref{fig:fig4}b we plot Eq.~(\ref{eq:avg_sig}) as a function of  $\varrho$ and $\delta r$. Black line shows the set of parameters resulting in $\langle \sigma \rangle \! = \! \sigma_0 = 8\mu$eV. Close to this line we mark four points which are used to generate the PSDs in Fig.~\ref{fig:fig5}.

In Fig.~\ref{fig:fig4} we also prove that $\text{Var}(\sigma)$ does not depend visibly on $\rho$, see an almost horizontal red line in Fig.~\ref{fig:fig4}). Consequently, in the relevant range of $\rho$ it is given by
    $\sqrt{\text{Var}(\sigma)} \approx a_{\text{std},0}\delta r,$
where $a_{\text{std},0} \approx 20$ $\mu\text{eV}$/nm.  This Var$(\sigma)$ is related to variability between statistically equivalent devices, which determines the reproducibility of QD properties and thus the yield of usable devices.\cite{Zwerver2022}

Finally, we use the above analysis and the obtained couplings $\{\eta_n\}$ to generate the noise PSDs, defined via Eq.~\eqref{eq:s_tls}. We concentrate on pairs ($\rho,\delta r$) that reproduce $\langle \sigma \rangle \approx 8$ $\mu$eV (the points in Fig.~\ref{fig:fig4}b).
 In Fig.~\ref{fig:fig5} we plot 10 realizations of the PSD and compare them with a  model $1/f$ noise spectrum with $S_1 \!=\! 1 \mu$eV$/$Hz (dashed lines) for each point marked in Fig.~\ref{fig:fig4}b. 
 Unsurprisingly, for all the sets we recover the PSD at 1 Hz $\approx \! 1$ $\mu$eV/Hz. However, for smallest $\rho = 5 \times 10^{9}$cm$^{-2}$ ($\mathrm{fpd} \! \approx \! 1.5$) and largest considered $\delta r \! =\! 0.2\,\mathrm{nm}$, in the uppermost panel of Fig.~\ref{fig:fig4} we see very strong effects of discrete number of TLFs, reflected in a locally Lorentzian PSD. While such strong single-TLF effects were seen in some experiments, \cite{kafanovChargeNoiseSingleelectron2008,Struck_NPJQI20,Connors_PRB19,Yoneda22} the PSD recently observed in Ref.~\onlinecite{Rojas23} is more closely matched by the results of the second panel, where $\rho\! \approx \! 10^{10}\,$cm$^{-2}$ and $\mathrm{fpd} \approx 2.7$, for which flat regions of PSD are less likely. The third panel ($\mathrm{fpd} \approx 6$) typically shows $S(f) \! \sim \! 1/f^\alpha$ with non-constant exponent $\alpha<2$, and resembles other experimentally measured PSDs.\cite{Struck_NPJQI20,Yoneda22, Connors_NC22} Finally, in the fourth panel, we have results with $\alpha \approx 1$, i.e.~spectra close to the ideal $1/f$ behavior. 

Summarizing, we have calculated electrostatic couplings between a quantum dot defined in realistic Si/SiGe nanostructure, and the charges trapped in the oxide above the semiconductor layer. Starting with reasonable initial ranges parameters for isotropic displacements of these charges, $\delta r$ and their densities, $\rho$, we have identified a set of $(\rho,\delta r)$ pairs, for which in the typically observed amplitude of $1/f$ charge noise at $1$ Hz, and its total power, can be reproduced. Deviations from ideal $1/f$ shape of charge noise PSDs  measured in various Si/SiGe structures\cite{kafanovChargeNoiseSingleelectron2008,Struck_NPJQI20,Connors_PRB19,Rojas23,Struck_NPJQI20,Yoneda22,Connors_NC22} can be reproduced with $\rho \geq 10^{10}$ cm$^{-2}$ and $\delta r \! \leq \! 0.15\,\mathrm{nm}$.
Note that these numbers might differ from actual values of $\rho$ and $\delta r$, as the QDs that work well as spin qubits could be subjected to selection bias, i.e.~focusing on QDs with lower-than typical amplitude of noise.
 The ranges of parameters found here to be consistent with experiments can be further narrowed down if more information on noise in a given nanostructure - for example on cross-correlations of noises in two QDs \cite{Yoneda22} - is available. Note that a recent paper\cite{Rojas23} shows that those correlations can be obtained in a model with $\rho$ similar to the one obtained here. Using the model and the parameter ranges identified here to calculation of PSD of spin splitting noise for a given direction of magnetic field gradient, is left for future work.

\begin{acknowledgments} 
This work was supported by funds from PRELUDIUM grant of the Polish National
Science Centre (NCN), Grant No. 2021/41/N/ST3/02758.
\end{acknowledgments} 

\section*{AUTHOR DECLARATIONS}
\subsection*{Conflict of interest}
The authors have no conflicts to disclose.

\section*{Data Availability Statement}
The data and code that support the findings of this study are available in the dedicated repository \cite{repo}.

\bibliographystyle{apsrev4-1}
\bibliography{literature}

\begin{thebibliography}{69}%
\makeatletter
\providecommand \@ifxundefined [1]{%
 \@ifx{#1\undefined}
}%
\providecommand \@ifnum [1]{%
 \ifnum #1\expandafter \@firstoftwo
 \else \expandafter \@secondoftwo
 \fi
}%
\providecommand \@ifx [1]{%
 \ifx #1\expandafter \@firstoftwo
 \else \expandafter \@secondoftwo
 \fi
}%
\providecommand \natexlab [1]{#1}%
\providecommand \enquote  [1]{``#1''}%
\providecommand \bibnamefont  [1]{#1}%
\providecommand \bibfnamefont [1]{#1}%
\providecommand \citenamefont [1]{#1}%
\providecommand \href@noop [0]{\@secondoftwo}%
\providecommand \href [0]{\begingroup \@sanitize@url \@href}%
\providecommand \@href[1]{\@@startlink{#1}\@@href}%
\providecommand \@@href[1]{\endgroup#1\@@endlink}%
\providecommand \@sanitize@url [0]{\catcode `\\12\catcode `\$12\catcode
  `\&12\catcode `\#12\catcode `\^12\catcode `\_12\catcode `\%12\relax}%
\providecommand \@@startlink[1]{}%
\providecommand \@@endlink[0]{}%
\providecommand \url  [0]{\begingroup\@sanitize@url \@url }%
\providecommand \@url [1]{\endgroup\@href {#1}{\urlprefix }}%
\providecommand \urlprefix  [0]{URL }%
\providecommand \Eprint [0]{\href }%
\providecommand \doibase [0]{http://dx.doi.org/}%
\providecommand \selectlanguage [0]{\@gobble}%
\providecommand \bibinfo  [0]{\@secondoftwo}%
\providecommand \bibfield  [0]{\@secondoftwo}%
\providecommand \translation [1]{[#1]}%
\providecommand \BibitemOpen [0]{}%
\providecommand \bibitemStop [0]{}%
\providecommand \bibitemNoStop [0]{.\EOS\space}%
\providecommand \EOS [0]{\spacefactor3000\relax}%
\providecommand \BibitemShut  [1]{\csname bibitem#1\endcsname}%
\let\auto@bib@innerbib\@empty
\bibitem [{\citenamefont {Dutta}\ and\ \citenamefont
  {Horn}(1981)}]{duttaLowfrequencyFluctuationsSolids1981}%
  \BibitemOpen
  \bibfield  {author} {\bibinfo {author} {\bibfnamefont {P.}~\bibnamefont
  {Dutta}}\ and\ \bibinfo {author} {\bibfnamefont {P.~M.}\ \bibnamefont
  {Horn}},\ }\href {\doibase 10.1103/RevModPhys.53.497} {\bibfield  {journal}
  {\bibinfo  {journal} {Reviews of Modern Physics}\ }\textbf {\bibinfo {volume}
  {53}},\ \bibinfo {pages} {497} (\bibinfo {year} {1981})}\BibitemShut
  {NoStop}%
\bibitem [{\citenamefont {Paladino}\ \emph {et~al.}(2014)\citenamefont
  {Paladino}, \citenamefont {Galperin}, \citenamefont {Falci},\ and\
  \citenamefont {Altshuler}}]{Paladino_RMP14}%
  \BibitemOpen
  \bibfield  {author} {\bibinfo {author} {\bibfnamefont {E.}~\bibnamefont
  {Paladino}}, \bibinfo {author} {\bibfnamefont {Y.~M.}\ \bibnamefont
  {Galperin}}, \bibinfo {author} {\bibfnamefont {G.}~\bibnamefont {Falci}}, \
  and\ \bibinfo {author} {\bibfnamefont {B.~L.}\ \bibnamefont {Altshuler}},\
  }\href {\doibase 10.1103/RevModPhys.86.361} {\bibfield  {journal} {\bibinfo
  {journal} {Rev. Mod. Phys.}\ }\textbf {\bibinfo {volume} {86}},\ \bibinfo
  {pages} {361} (\bibinfo {year} {2014})}\BibitemShut {NoStop}%
\bibitem [{\citenamefont
  {Fleetwood}(2015)}]{fleetwoodNoiseDefectsMicroelectronic2015}%
  \BibitemOpen
  \bibfield  {author} {\bibinfo {author} {\bibfnamefont {D.~M.}\ \bibnamefont
  {Fleetwood}},\ }\href {\doibase 10.1109/TNS.2015.2405852} {\bibfield
  {journal} {\bibinfo  {journal} {IEEE Transactions on Nuclear Science}\
  }\textbf {\bibinfo {volume} {62}},\ \bibinfo {pages} {1462} (\bibinfo {year}
  {2015})}\BibitemShut {NoStop}%
\bibitem [{\citenamefont {Islam}\ \emph {et~al.}(2022)\citenamefont {Islam},
  \citenamefont {Shamim},\ and\ \citenamefont
  {Ghosh}}]{islamBenchmarkingNoiseDephasing2022}%
  \BibitemOpen
  \bibfield  {author} {\bibinfo {author} {\bibfnamefont {S.}~\bibnamefont
  {Islam}}, \bibinfo {author} {\bibfnamefont {S.}~\bibnamefont {Shamim}}, \
  and\ \bibinfo {author} {\bibfnamefont {A.}~\bibnamefont {Ghosh}},\ }\href
  {\doibase 10.1002/adma.202109671} {\bibfield  {journal} {\bibinfo  {journal}
  {Advanced Materials}\ ,\ \bibinfo {pages} {2109671}} (\bibinfo {year}
  {2022})}\BibitemShut {NoStop}%
\bibitem [{\citenamefont {Wuetz}\ \emph {et~al.}(2023)\citenamefont {Wuetz},
  \citenamefont {Esposti}, \citenamefont {Zwerver}, \citenamefont {Amitonov},
  \citenamefont {Botifoll}, \citenamefont {Arbiol}, \citenamefont {Sammak},
  \citenamefont {Vandersypen}, \citenamefont {Russ},\ and\ \citenamefont
  {Scappucci}}]{Wuetz_NC23}%
  \BibitemOpen
  \bibfield  {author} {\bibinfo {author} {\bibfnamefont {B.~P.}\ \bibnamefont
  {Wuetz}}, \bibinfo {author} {\bibfnamefont {D.~D.}\ \bibnamefont {Esposti}},
  \bibinfo {author} {\bibfnamefont {A.~M.~J.}\ \bibnamefont {Zwerver}},
  \bibinfo {author} {\bibfnamefont {S.~V.}\ \bibnamefont {Amitonov}}, \bibinfo
  {author} {\bibfnamefont {M.}~\bibnamefont {Botifoll}}, \bibinfo {author}
  {\bibfnamefont {J.}~\bibnamefont {Arbiol}}, \bibinfo {author} {\bibfnamefont
  {A.}~\bibnamefont {Sammak}}, \bibinfo {author} {\bibfnamefont {L.~M.~K.}\
  \bibnamefont {Vandersypen}}, \bibinfo {author} {\bibfnamefont
  {M.}~\bibnamefont {Russ}}, \ and\ \bibinfo {author} {\bibfnamefont
  {G.}~\bibnamefont {Scappucci}},\ }\href {\doibase 10.1038/s41467-023-36951-w}
  {\bibfield  {journal} {\bibinfo  {journal} {Nat.~Comm.}\ }\textbf {\bibinfo
  {volume} {14}},\ \bibinfo {pages} {1385} (\bibinfo {year}
  {2023})}\BibitemShut {NoStop}%
\bibitem [{\citenamefont {Sakamoto}\ \emph {et~al.}(1995)\citenamefont
  {Sakamoto}, \citenamefont {Nakamura},\ and\ \citenamefont
  {Nakamura}}]{sakamoto_nakamura_nakamura_1995}%
  \BibitemOpen
  \bibfield  {author} {\bibinfo {author} {\bibfnamefont {T.}~\bibnamefont
  {Sakamoto}}, \bibinfo {author} {\bibfnamefont {Y.}~\bibnamefont {Nakamura}},
  \ and\ \bibinfo {author} {\bibfnamefont {K.}~\bibnamefont {Nakamura}},\
  }\href {\doibase 10.1063/1.115109} {\bibfield  {journal} {\bibinfo  {journal}
  {Applied Physics Letters}\ }\textbf {\bibinfo {volume} {67}},\ \bibinfo
  {pages} {2220–2222} (\bibinfo {year} {1995})}\BibitemShut {NoStop}%
\bibitem [{\citenamefont {Liefrink}\ \emph {et~al.}(1994)\citenamefont
  {Liefrink}, \citenamefont {Dijkhuis},\ and\ \citenamefont
  {Houten}}]{liefrink_dijkhuis_houten_1994}%
  \BibitemOpen
  \bibfield  {author} {\bibinfo {author} {\bibfnamefont {F.}~\bibnamefont
  {Liefrink}}, \bibinfo {author} {\bibfnamefont {J.~I.}\ \bibnamefont
  {Dijkhuis}}, \ and\ \bibinfo {author} {\bibfnamefont {H.~v.}\ \bibnamefont
  {Houten}},\ }\href {\doibase https://doi.org/10.1088/0268-1242/9/12/003}
  {\bibfield  {journal} {\bibinfo  {journal} {Semiconductor Science and
  Technology}\ }\textbf {\bibinfo {volume} {9}},\ \bibinfo {pages}
  {2178–2189} (\bibinfo {year} {1994})}\BibitemShut {NoStop}%
\bibitem [{\citenamefont {Ramon}\ and\ \citenamefont
  {Hu}(2010)}]{ramonDecoherenceSpinQubits2010}%
  \BibitemOpen
  \bibfield  {author} {\bibinfo {author} {\bibfnamefont {G.}~\bibnamefont
  {Ramon}}\ and\ \bibinfo {author} {\bibfnamefont {X.}~\bibnamefont {Hu}},\
  }\href {\doibase 10.1103/PhysRevB.81.045304} {\bibfield  {journal} {\bibinfo
  {journal} {Physical Review B}\ }\textbf {\bibinfo {volume} {81}},\ \bibinfo
  {pages} {045304} (\bibinfo {year} {2010})}\BibitemShut {NoStop}%
\bibitem [{\citenamefont {Bermeister}\ \emph {et~al.}(2014)\citenamefont
  {Bermeister}, \citenamefont {Keith},\ and\ \citenamefont
  {Culcer}}]{bermeisterChargeNoiseSpinorbit2014}%
  \BibitemOpen
  \bibfield  {author} {\bibinfo {author} {\bibfnamefont {A.}~\bibnamefont
  {Bermeister}}, \bibinfo {author} {\bibfnamefont {D.}~\bibnamefont {Keith}}, \
  and\ \bibinfo {author} {\bibfnamefont {D.}~\bibnamefont {Culcer}},\ }\href
  {\doibase 10.1063/1.4901162} {\bibfield  {journal} {\bibinfo  {journal}
  {Applied Physics Letters}\ }\textbf {\bibinfo {volume} {105}},\ \bibinfo
  {pages} {192102} (\bibinfo {year} {2014})}\BibitemShut {NoStop}%
\bibitem [{\citenamefont {Hanson}\ \emph {et~al.}(2007)\citenamefont {Hanson},
  \citenamefont {Kouwenhoven}, \citenamefont {Petta}, \citenamefont {Tarucha},\
  and\ \citenamefont {Vandersypen}}]{Hanson_RMP07}%
  \BibitemOpen
  \bibfield  {author} {\bibinfo {author} {\bibfnamefont {R.}~\bibnamefont
  {Hanson}}, \bibinfo {author} {\bibfnamefont {L.~P.}\ \bibnamefont
  {Kouwenhoven}}, \bibinfo {author} {\bibfnamefont {J.~R.}\ \bibnamefont
  {Petta}}, \bibinfo {author} {\bibfnamefont {S.}~\bibnamefont {Tarucha}}, \
  and\ \bibinfo {author} {\bibfnamefont {L.~M.~K.}\ \bibnamefont
  {Vandersypen}},\ }\href {\doibase 10.1103/RevModPhys.79.1217} {\bibfield
  {journal} {\bibinfo  {journal} {Reviews of Modern Physics}\ }\textbf
  {\bibinfo {volume} {79}},\ \bibinfo {pages} {1217} (\bibinfo {year}
  {2007})}\BibitemShut {NoStop}%
\bibitem [{\citenamefont {Burkard}\ \emph {et~al.}(2021)\citenamefont
  {Burkard}, \citenamefont {Ladd}, \citenamefont {Nichol}, \citenamefont
  {Pan},\ and\ \citenamefont {Petta}}]{Burkard21}%
  \BibitemOpen
  \bibfield  {author} {\bibinfo {author} {\bibfnamefont {G.}~\bibnamefont
  {Burkard}}, \bibinfo {author} {\bibfnamefont {T.~D.}\ \bibnamefont {Ladd}},
  \bibinfo {author} {\bibfnamefont {J.~M.}\ \bibnamefont {Nichol}}, \bibinfo
  {author} {\bibfnamefont {A.}~\bibnamefont {Pan}}, \ and\ \bibinfo {author}
  {\bibfnamefont {J.~R.}\ \bibnamefont {Petta}},\ }\href
  {http://arxiv.org/abs/2112.08863} {\bibfield  {journal} {\bibinfo  {journal}
  {arXiv:2112.08863}\ } (\bibinfo {year} {2021})}\BibitemShut {NoStop}%
\bibitem [{\citenamefont {Philips}\ \emph {et~al.}(2022)\citenamefont
  {Philips}, \citenamefont {Madzik}, \citenamefont {Amitonov}, \citenamefont
  {{de Snoo}}, \citenamefont {Russ}, \citenamefont {Kalhor}, \citenamefont
  {Volk}, \citenamefont {Lawrie}, \citenamefont {Brousse}, \citenamefont
  {Tryputen}, \citenamefont {Wuetz}, \citenamefont {Sammak}, \citenamefont
  {Veldhorst}, \citenamefont {Scappucci},\ and\ \citenamefont
  {Vandersypen}}]{philipsUniversalControlSixqubit2022}%
  \BibitemOpen
  \bibfield  {author} {\bibinfo {author} {\bibfnamefont {S.~G.~J.}\
  \bibnamefont {Philips}}, \bibinfo {author} {\bibfnamefont {M.~T.}\
  \bibnamefont {Madzik}}, \bibinfo {author} {\bibfnamefont {S.~V.}\
  \bibnamefont {Amitonov}}, \bibinfo {author} {\bibfnamefont {S.~L.}\
  \bibnamefont {{de Snoo}}}, \bibinfo {author} {\bibfnamefont {M.}~\bibnamefont
  {Russ}}, \bibinfo {author} {\bibfnamefont {N.}~\bibnamefont {Kalhor}},
  \bibinfo {author} {\bibfnamefont {C.}~\bibnamefont {Volk}}, \bibinfo {author}
  {\bibfnamefont {W.~I.~L.}\ \bibnamefont {Lawrie}}, \bibinfo {author}
  {\bibfnamefont {D.}~\bibnamefont {Brousse}}, \bibinfo {author} {\bibfnamefont
  {L.}~\bibnamefont {Tryputen}}, \bibinfo {author} {\bibfnamefont {B.~P.}\
  \bibnamefont {Wuetz}}, \bibinfo {author} {\bibfnamefont {A.}~\bibnamefont
  {Sammak}}, \bibinfo {author} {\bibfnamefont {M.}~\bibnamefont {Veldhorst}},
  \bibinfo {author} {\bibfnamefont {G.}~\bibnamefont {Scappucci}}, \ and\
  \bibinfo {author} {\bibfnamefont {L.~M.~K.}\ \bibnamefont {Vandersypen}},\
  }\href {\doibase 10.1038/s41586-022-05117-x} {\bibfield  {journal} {\bibinfo
  {journal} {Nature}\ }\textbf {\bibinfo {volume} {609}},\ \bibinfo {pages}
  {919} (\bibinfo {year} {2022})}\BibitemShut {NoStop}%
\bibitem [{\citenamefont {Borsoi}\ \emph {et~al.}(2022)\citenamefont {Borsoi},
  \citenamefont {Hendrickx}, \citenamefont {John}, \citenamefont {Motz},
  \citenamefont {van Riggelen}, \citenamefont {Sammak}, \citenamefont {{de
  Snoo}}, \citenamefont {Scappucci},\ and\ \citenamefont
  {Veldhorst}}]{Borsoi22}%
  \BibitemOpen
  \bibfield  {author} {\bibinfo {author} {\bibfnamefont {F.}~\bibnamefont
  {Borsoi}}, \bibinfo {author} {\bibfnamefont {N.~W.}\ \bibnamefont
  {Hendrickx}}, \bibinfo {author} {\bibfnamefont {V.}~\bibnamefont {John}},
  \bibinfo {author} {\bibfnamefont {S.}~\bibnamefont {Motz}}, \bibinfo {author}
  {\bibfnamefont {F.}~\bibnamefont {van Riggelen}}, \bibinfo {author}
  {\bibfnamefont {A.}~\bibnamefont {Sammak}}, \bibinfo {author} {\bibfnamefont
  {S.~L.}\ \bibnamefont {{de Snoo}}}, \bibinfo {author} {\bibfnamefont
  {G.}~\bibnamefont {Scappucci}}, \ and\ \bibinfo {author} {\bibfnamefont
  {M.}~\bibnamefont {Veldhorst}},\ }\href {http://arxiv.org/abs/2209.06609}
  {\bibfield  {journal} {\bibinfo  {journal} {arXiv:2209.06609}\ } (\bibinfo
  {year} {2022})}\BibitemShut {NoStop}%
\bibitem [{\citenamefont {Huang}\ \emph {et~al.}(2018)\citenamefont {Huang},
  \citenamefont {Zimmerman},\ and\ \citenamefont
  {Bryant}}]{huangSpinDecoherenceTwoqubit2018}%
  \BibitemOpen
  \bibfield  {author} {\bibinfo {author} {\bibfnamefont {P.}~\bibnamefont
  {Huang}}, \bibinfo {author} {\bibfnamefont {N.~M.}\ \bibnamefont
  {Zimmerman}}, \ and\ \bibinfo {author} {\bibfnamefont {G.~W.}\ \bibnamefont
  {Bryant}},\ }\href {\doibase 10.1038/s41534-018-0112-0} {\bibfield  {journal}
  {\bibinfo  {journal} {npj Quantum Information}\ }\textbf {\bibinfo {volume}
  {4}},\ \bibinfo {pages} {62} (\bibinfo {year} {2018})}\BibitemShut {NoStop}%
\bibitem [{\citenamefont {Veldhorst}\ \emph {et~al.}(2015)\citenamefont
  {Veldhorst}, \citenamefont {Yang}, \citenamefont {Hwang}, \citenamefont
  {Huang}, \citenamefont {Dehollain}, \citenamefont {Muhonen}, \citenamefont
  {Simmons}, \citenamefont {Laucht}, \citenamefont {Hudson}, \citenamefont
  {Itoh}, \citenamefont {Morello},\ and\ \citenamefont
  {Dzurak}}]{veldhorstTwoqubitLogicGate2015a}%
  \BibitemOpen
  \bibfield  {author} {\bibinfo {author} {\bibfnamefont {M.}~\bibnamefont
  {Veldhorst}}, \bibinfo {author} {\bibfnamefont {C.~H.}\ \bibnamefont {Yang}},
  \bibinfo {author} {\bibfnamefont {J.~C.~C.}\ \bibnamefont {Hwang}}, \bibinfo
  {author} {\bibfnamefont {W.}~\bibnamefont {Huang}}, \bibinfo {author}
  {\bibfnamefont {J.~P.}\ \bibnamefont {Dehollain}}, \bibinfo {author}
  {\bibfnamefont {J.~T.}\ \bibnamefont {Muhonen}}, \bibinfo {author}
  {\bibfnamefont {S.}~\bibnamefont {Simmons}}, \bibinfo {author} {\bibfnamefont
  {A.}~\bibnamefont {Laucht}}, \bibinfo {author} {\bibfnamefont {F.~E.}\
  \bibnamefont {Hudson}}, \bibinfo {author} {\bibfnamefont {K.~M.}\
  \bibnamefont {Itoh}}, \bibinfo {author} {\bibfnamefont {A.}~\bibnamefont
  {Morello}}, \ and\ \bibinfo {author} {\bibfnamefont {A.~S.}\ \bibnamefont
  {Dzurak}},\ }\href {\doibase 10.1038/nature15263} {\bibfield  {journal}
  {\bibinfo  {journal} {Nature}\ }\textbf {\bibinfo {volume} {526}},\ \bibinfo
  {pages} {410} (\bibinfo {year} {2015})}\BibitemShut {NoStop}%
\bibitem [{\citenamefont {Zajac}\ \emph {et~al.}(2018)\citenamefont {Zajac},
  \citenamefont {Sigillito}, \citenamefont {Russ}, \citenamefont {Borjans},
  \citenamefont {Taylor}, \citenamefont {Burkard},\ and\ \citenamefont {{J. R.
  Petta}}}]{ZajacScience18}%
  \BibitemOpen
  \bibfield  {author} {\bibinfo {author} {\bibfnamefont {D.~M.}\ \bibnamefont
  {Zajac}}, \bibinfo {author} {\bibfnamefont {A.~J.}\ \bibnamefont
  {Sigillito}}, \bibinfo {author} {\bibfnamefont {M.}~\bibnamefont {Russ}},
  \bibinfo {author} {\bibfnamefont {F.}~\bibnamefont {Borjans}}, \bibinfo
  {author} {\bibfnamefont {J.~M.}\ \bibnamefont {Taylor}}, \bibinfo {author}
  {\bibfnamefont {G.}~\bibnamefont {Burkard}}, \ and\ \bibinfo {author}
  {\bibnamefont {{J. R. Petta}}},\ }\href {\doibase 10.1126/science.aao5965}
  {\bibfield  {journal} {\bibinfo  {journal} {Science (New York, N.Y.)}\
  }\textbf {\bibinfo {volume} {359}},\ \bibinfo {pages} {439} (\bibinfo {year}
  {2018})}\BibitemShut {NoStop}%
\bibitem [{\citenamefont {Watson}\ \emph {et~al.}(2018)\citenamefont {Watson},
  \citenamefont {Philips}, \citenamefont {Kawakami}, \citenamefont {Ward},
  \citenamefont {Scarlino}, \citenamefont {Veldhorst}, \citenamefont {Savage},
  \citenamefont {Lagally}, \citenamefont {Friesen}, \citenamefont
  {Coppersmith}, \citenamefont {Eriksson},\ and\ \citenamefont
  {Vandersypen}}]{watsonProgrammableTwoqubitQuantum2018b}%
  \BibitemOpen
  \bibfield  {author} {\bibinfo {author} {\bibfnamefont {T.~F.}\ \bibnamefont
  {Watson}}, \bibinfo {author} {\bibfnamefont {S.~G.~J.}\ \bibnamefont
  {Philips}}, \bibinfo {author} {\bibfnamefont {E.}~\bibnamefont {Kawakami}},
  \bibinfo {author} {\bibfnamefont {D.~R.}\ \bibnamefont {Ward}}, \bibinfo
  {author} {\bibfnamefont {P.}~\bibnamefont {Scarlino}}, \bibinfo {author}
  {\bibfnamefont {M.}~\bibnamefont {Veldhorst}}, \bibinfo {author}
  {\bibfnamefont {D.~E.}\ \bibnamefont {Savage}}, \bibinfo {author}
  {\bibfnamefont {M.~G.}\ \bibnamefont {Lagally}}, \bibinfo {author}
  {\bibfnamefont {M.}~\bibnamefont {Friesen}}, \bibinfo {author} {\bibfnamefont
  {S.~N.}\ \bibnamefont {Coppersmith}}, \bibinfo {author} {\bibfnamefont
  {M.~A.}\ \bibnamefont {Eriksson}}, \ and\ \bibinfo {author} {\bibfnamefont
  {L.~M.~K.}\ \bibnamefont {Vandersypen}},\ }\href {\doibase
  10.1038/nature25766} {\bibfield  {journal} {\bibinfo  {journal} {Nature}\
  }\textbf {\bibinfo {volume} {555}},\ \bibinfo {pages} {633} (\bibinfo {year}
  {2018})}\BibitemShut {NoStop}%
\bibitem [{\citenamefont {Coish}\ and\ \citenamefont
  {Loss}(2005)}]{Coish_PRB05}%
  \BibitemOpen
  \bibfield  {author} {\bibinfo {author} {\bibfnamefont {W.~A.}\ \bibnamefont
  {Coish}}\ and\ \bibinfo {author} {\bibfnamefont {D.}~\bibnamefont {Loss}},\
  }\href {\doibase 10.1103/PhysRevB.72.125337} {\bibfield  {journal} {\bibinfo
  {journal} {Phys. Rev. B}\ }\textbf {\bibinfo {volume} {72}},\ \bibinfo
  {pages} {125337} (\bibinfo {year} {2005})}\BibitemShut {NoStop}%
\bibitem [{\citenamefont {Hu}\ and\ \citenamefont {{Das
  Sarma}}(2006)}]{Hu_PRL06}%
  \BibitemOpen
  \bibfield  {author} {\bibinfo {author} {\bibfnamefont {X.}~\bibnamefont
  {Hu}}\ and\ \bibinfo {author} {\bibfnamefont {S.}~\bibnamefont {{Das
  Sarma}}},\ }\href {\doibase 10.1103/PhysRevLett.96.100501} {\bibfield
  {journal} {\bibinfo  {journal} {Phys. Rev. Lett.}\ }\textbf {\bibinfo
  {volume} {96}},\ \bibinfo {pages} {100501} (\bibinfo {year}
  {2006})}\BibitemShut {NoStop}%
\bibitem [{\citenamefont {Dial}\ \emph {et~al.}(2013)\citenamefont {Dial},
  \citenamefont {Shulman}, \citenamefont {Harvey}, \citenamefont {Bluhm},
  \citenamefont {Umansky},\ and\ \citenamefont {Yacoby}}]{Dial_PRL13}%
  \BibitemOpen
  \bibfield  {author} {\bibinfo {author} {\bibfnamefont {O.~E.}\ \bibnamefont
  {Dial}}, \bibinfo {author} {\bibfnamefont {M.~D.}\ \bibnamefont {Shulman}},
  \bibinfo {author} {\bibfnamefont {S.~P.}\ \bibnamefont {Harvey}}, \bibinfo
  {author} {\bibfnamefont {H.}~\bibnamefont {Bluhm}}, \bibinfo {author}
  {\bibfnamefont {V.}~\bibnamefont {Umansky}}, \ and\ \bibinfo {author}
  {\bibfnamefont {A.}~\bibnamefont {Yacoby}},\ }\href {\doibase
  10.1103/PhysRevLett.110.146804} {\bibfield  {journal} {\bibinfo  {journal}
  {Phys. Rev. Lett.}\ }\textbf {\bibinfo {volume} {110}},\ \bibinfo {pages}
  {146804} (\bibinfo {year} {2013})}\BibitemShut {NoStop}%
\bibitem [{\citenamefont {Takeda}\ \emph {et~al.}(2020)\citenamefont {Takeda},
  \citenamefont {Noiri}, \citenamefont {Yoneda}, \citenamefont {Nakajima},\
  and\ \citenamefont {Tarucha}}]{TakedaPRL2020}%
  \BibitemOpen
  \bibfield  {author} {\bibinfo {author} {\bibfnamefont {K.}~\bibnamefont
  {Takeda}}, \bibinfo {author} {\bibfnamefont {A.}~\bibnamefont {Noiri}},
  \bibinfo {author} {\bibfnamefont {J.}~\bibnamefont {Yoneda}}, \bibinfo
  {author} {\bibfnamefont {T.}~\bibnamefont {Nakajima}}, \ and\ \bibinfo
  {author} {\bibfnamefont {S.}~\bibnamefont {Tarucha}},\ }\href {\doibase
  10.1103/PhysRevLett.124.117701} {\bibfield  {journal} {\bibinfo  {journal}
  {Physical Review Letters}\ }\textbf {\bibinfo {volume} {124}},\ \bibinfo
  {pages} {117701} (\bibinfo {year} {2020})}\BibitemShut {NoStop}%
\bibitem [{\citenamefont {Connors}\ \emph {et~al.}(2022)\citenamefont
  {Connors}, \citenamefont {Nelson}, \citenamefont {Edge},\ and\ \citenamefont
  {Nichol}}]{Connors_NC22}%
  \BibitemOpen
  \bibfield  {author} {\bibinfo {author} {\bibfnamefont {E.~J.}\ \bibnamefont
  {Connors}}, \bibinfo {author} {\bibfnamefont {J.}~\bibnamefont {Nelson}},
  \bibinfo {author} {\bibfnamefont {L.~F.}\ \bibnamefont {Edge}}, \ and\
  \bibinfo {author} {\bibfnamefont {J.~M.}\ \bibnamefont {Nichol}},\ }\href
  {\doibase 10.1038/s41467-022-28519-x} {\bibfield  {journal} {\bibinfo
  {journal} {Nat.~Comm.}\ }\textbf {\bibinfo {volume} {13}},\ \bibinfo {pages}
  {940} (\bibinfo {year} {2022})}\BibitemShut {NoStop}%
\bibitem [{\citenamefont {Chan}\ \emph {et~al.}(2018)\citenamefont {Chan},
  \citenamefont {Huang}, \citenamefont {Yang}, \citenamefont {Hwang},
  \citenamefont {Hensen}, \citenamefont {Tanttu}, \citenamefont {Hudson},
  \citenamefont {Itoh}, \citenamefont {Laucht}, \citenamefont {Morello},\ and\
  \citenamefont {Dzurak}}]{Chan_PRAP18}%
  \BibitemOpen
  \bibfield  {author} {\bibinfo {author} {\bibfnamefont {K.~W.}\ \bibnamefont
  {Chan}}, \bibinfo {author} {\bibfnamefont {W.}~\bibnamefont {Huang}},
  \bibinfo {author} {\bibfnamefont {C.~H.}\ \bibnamefont {Yang}}, \bibinfo
  {author} {\bibfnamefont {J.~C.}\ \bibnamefont {Hwang}}, \bibinfo {author}
  {\bibfnamefont {B.}~\bibnamefont {Hensen}}, \bibinfo {author} {\bibfnamefont
  {T.}~\bibnamefont {Tanttu}}, \bibinfo {author} {\bibfnamefont {F.~E.}\
  \bibnamefont {Hudson}}, \bibinfo {author} {\bibfnamefont {K.~M.}\
  \bibnamefont {Itoh}}, \bibinfo {author} {\bibfnamefont {A.}~\bibnamefont
  {Laucht}}, \bibinfo {author} {\bibfnamefont {A.}~\bibnamefont {Morello}}, \
  and\ \bibinfo {author} {\bibfnamefont {A.~S.}\ \bibnamefont {Dzurak}},\
  }\href {\doibase 10.1103/PhysRevApplied.10.044017} {\bibfield  {journal}
  {\bibinfo  {journal} {Physical Review Applied}\ }\textbf {\bibinfo {volume}
  {10}},\ \bibinfo {pages} {1} (\bibinfo {year} {2018})}\BibitemShut {NoStop}%
\bibitem [{\citenamefont {Qi}\ \emph {et~al.}(2017)\citenamefont {Qi},
  \citenamefont {Wu}, \citenamefont {Ward}, \citenamefont {Prance},
  \citenamefont {Kim}, \citenamefont {Gamble}, \citenamefont {Mohr},
  \citenamefont {Shi}, \citenamefont {Savage}, \citenamefont {Lagally},
  \citenamefont {Eriksson}, \citenamefont {Friesen}, \citenamefont
  {Coppersmith},\ and\ \citenamefont {Vavilov}}]{Qi_PRB17}%
  \BibitemOpen
  \bibfield  {author} {\bibinfo {author} {\bibfnamefont {Z.}~\bibnamefont
  {Qi}}, \bibinfo {author} {\bibfnamefont {X.}~\bibnamefont {Wu}}, \bibinfo
  {author} {\bibfnamefont {D.~R.}\ \bibnamefont {Ward}}, \bibinfo {author}
  {\bibfnamefont {J.~R.}\ \bibnamefont {Prance}}, \bibinfo {author}
  {\bibfnamefont {D.}~\bibnamefont {Kim}}, \bibinfo {author} {\bibfnamefont
  {J.~K.}\ \bibnamefont {Gamble}}, \bibinfo {author} {\bibfnamefont {R.~T.}\
  \bibnamefont {Mohr}}, \bibinfo {author} {\bibfnamefont {Z.}~\bibnamefont
  {Shi}}, \bibinfo {author} {\bibfnamefont {D.~E.}\ \bibnamefont {Savage}},
  \bibinfo {author} {\bibfnamefont {M.~G.}\ \bibnamefont {Lagally}}, \bibinfo
  {author} {\bibfnamefont {M.~A.}\ \bibnamefont {Eriksson}}, \bibinfo {author}
  {\bibfnamefont {M.}~\bibnamefont {Friesen}}, \bibinfo {author} {\bibfnamefont
  {S.~N.}\ \bibnamefont {Coppersmith}}, \ and\ \bibinfo {author} {\bibfnamefont
  {M.~G.}\ \bibnamefont {Vavilov}},\ }\href {\doibase
  10.1103/PhysRevB.96.115305} {\bibfield  {journal} {\bibinfo  {journal}
  {Phys.~Rev.~B}\ }\textbf {\bibinfo {volume} {96}},\ \bibinfo {pages} {115305}
  (\bibinfo {year} {2017})}\BibitemShut {NoStop}%
\bibitem [{\citenamefont {Culcer}\ and\ \citenamefont
  {Zimmerman}(2013)}]{culcerDephasingSiSinglettriplet2013}%
  \BibitemOpen
  \bibfield  {author} {\bibinfo {author} {\bibfnamefont {D.}~\bibnamefont
  {Culcer}}\ and\ \bibinfo {author} {\bibfnamefont {N.~M.}\ \bibnamefont
  {Zimmerman}},\ }\href {\doibase 10.1063/1.4810911} {\bibfield  {journal}
  {\bibinfo  {journal} {Applied Physics Letters}\ }\textbf {\bibinfo {volume}
  {102}},\ \bibinfo {pages} {232108} (\bibinfo {year} {2013})}\BibitemShut
  {NoStop}%
\bibitem [{\citenamefont {Yoneda}\ \emph {et~al.}(2018)\citenamefont {Yoneda},
  \citenamefont {Takeda}, \citenamefont {Otsuka}, \citenamefont {Nakajima},
  \citenamefont {Delbecq}, \citenamefont {Allison}, \citenamefont {Honda},
  \citenamefont {Kodera}, \citenamefont {Oda}, \citenamefont {Hoshi},
  \citenamefont {Usami}, \citenamefont {Itoh},\ and\ \citenamefont
  {Tarucha}}]{Yoneda_NN18}%
  \BibitemOpen
  \bibfield  {author} {\bibinfo {author} {\bibfnamefont {J.}~\bibnamefont
  {Yoneda}}, \bibinfo {author} {\bibfnamefont {K.}~\bibnamefont {Takeda}},
  \bibinfo {author} {\bibfnamefont {T.}~\bibnamefont {Otsuka}}, \bibinfo
  {author} {\bibfnamefont {T.}~\bibnamefont {Nakajima}}, \bibinfo {author}
  {\bibfnamefont {M.~R.}\ \bibnamefont {Delbecq}}, \bibinfo {author}
  {\bibfnamefont {G.}~\bibnamefont {Allison}}, \bibinfo {author} {\bibfnamefont
  {T.}~\bibnamefont {Honda}}, \bibinfo {author} {\bibfnamefont
  {T.}~\bibnamefont {Kodera}}, \bibinfo {author} {\bibfnamefont
  {S.}~\bibnamefont {Oda}}, \bibinfo {author} {\bibfnamefont {Y.}~\bibnamefont
  {Hoshi}}, \bibinfo {author} {\bibfnamefont {N.}~\bibnamefont {Usami}},
  \bibinfo {author} {\bibfnamefont {K.~M.}\ \bibnamefont {Itoh}}, \ and\
  \bibinfo {author} {\bibfnamefont {S.}~\bibnamefont {Tarucha}},\ }\href
  {\doibase 10.1038/s41565-017-0014-x} {\bibfield  {journal} {\bibinfo
  {journal} {Nature Nanotechnology}\ }\textbf {\bibinfo {volume} {13}},\
  \bibinfo {pages} {102} (\bibinfo {year} {2018})}\BibitemShut {NoStop}%
\bibitem [{\citenamefont {Struck}\ \emph {et~al.}(2029)\citenamefont {Struck},
  \citenamefont {Hollmann}, \citenamefont {Schauer}, \citenamefont {Fedorets},
  \citenamefont {Schmidbauer}, \citenamefont {Sawano}, \citenamefont {Riemann},
  \citenamefont {Abrosimov}, \citenamefont {Cywi{\'n}ski}, \citenamefont
  {Bougeard},\ and\ \citenamefont {Schreiber}}]{Struck_NPJQI20}%
  \BibitemOpen
  \bibfield  {author} {\bibinfo {author} {\bibfnamefont {T.}~\bibnamefont
  {Struck}}, \bibinfo {author} {\bibfnamefont {A.}~\bibnamefont {Hollmann}},
  \bibinfo {author} {\bibfnamefont {F.}~\bibnamefont {Schauer}}, \bibinfo
  {author} {\bibfnamefont {O.}~\bibnamefont {Fedorets}}, \bibinfo {author}
  {\bibfnamefont {A.}~\bibnamefont {Schmidbauer}}, \bibinfo {author}
  {\bibfnamefont {K.}~\bibnamefont {Sawano}}, \bibinfo {author} {\bibfnamefont
  {H.}~\bibnamefont {Riemann}}, \bibinfo {author} {\bibfnamefont {N.~V.}\
  \bibnamefont {Abrosimov}}, \bibinfo {author} {\bibfnamefont
  {{\L}.}~\bibnamefont {Cywi{\'n}ski}}, \bibinfo {author} {\bibfnamefont
  {D.}~\bibnamefont {Bougeard}}, \ and\ \bibinfo {author} {\bibfnamefont
  {L.~R.}\ \bibnamefont {Schreiber}},\ }\href {\doibase
  10.1038/s41534-020-0276-2} {\bibfield  {journal} {\bibinfo  {journal} {npj
  Quantum Inf.}\ }\textbf {\bibinfo {volume} {6}},\ \bibinfo {pages} {40}
  (\bibinfo {year} {2029})}\BibitemShut {NoStop}%
\bibitem [{\citenamefont {Saraiva}\ \emph {et~al.}(2021)\citenamefont
  {Saraiva}, \citenamefont {Lim}, \citenamefont {Yang}, \citenamefont {Escott},
  \citenamefont {Laucht},\ and\ \citenamefont {Dzurak}}]{Saraiva_AFM21}%
  \BibitemOpen
  \bibfield  {author} {\bibinfo {author} {\bibfnamefont {A.}~\bibnamefont
  {Saraiva}}, \bibinfo {author} {\bibfnamefont {W.~H.}\ \bibnamefont {Lim}},
  \bibinfo {author} {\bibfnamefont {C.~H.}\ \bibnamefont {Yang}}, \bibinfo
  {author} {\bibfnamefont {C.~C.}\ \bibnamefont {Escott}}, \bibinfo {author}
  {\bibfnamefont {A.}~\bibnamefont {Laucht}}, \ and\ \bibinfo {author}
  {\bibfnamefont {A.~S.}\ \bibnamefont {Dzurak}},\ }\href {\doibase
  10.1002/adfm.202105488} {\bibfield  {journal} {\bibinfo  {journal}
  {Adv.~Funct.~Mater.}\ }\textbf {\bibinfo {volume} {32}},\ \bibinfo {pages}
  {2105488} (\bibinfo {year} {2021})}\BibitemShut {NoStop}%
\bibitem [{\citenamefont {Gilbert}\ \emph {et~al.}(2023)\citenamefont
  {Gilbert}, \citenamefont {Tanttu}, \citenamefont {Lim}, \citenamefont {Feng},
  \citenamefont {Huang}, \citenamefont {Cifuentes}, \citenamefont {Serrano},
  \citenamefont {Mai}, \citenamefont {Leon}, \citenamefont {Escott},
  \citenamefont {Itoh}, \citenamefont {Abrosimov}, \citenamefont {Pohl},
  \citenamefont {Thewalt}, \citenamefont {Hudson}, \citenamefont {Morello},
  \citenamefont {Laucht}, \citenamefont {Yang}, \citenamefont {Saraiva},\ and\
  \citenamefont
  {Dzurak}}]{gilbert_tanttu_lim_feng_huang_cifuentes_serrano_mai_leon_escott_etal_2023}%
  \BibitemOpen
  \bibfield  {author} {\bibinfo {author} {\bibfnamefont {W.}~\bibnamefont
  {Gilbert}}, \bibinfo {author} {\bibfnamefont {T.}~\bibnamefont {Tanttu}},
  \bibinfo {author} {\bibfnamefont {W.~H.}\ \bibnamefont {Lim}}, \bibinfo
  {author} {\bibfnamefont {M.}~\bibnamefont {Feng}}, \bibinfo {author}
  {\bibfnamefont {J.~Y.}\ \bibnamefont {Huang}}, \bibinfo {author}
  {\bibfnamefont {J.~D.}\ \bibnamefont {Cifuentes}}, \bibinfo {author}
  {\bibfnamefont {S.}~\bibnamefont {Serrano}}, \bibinfo {author} {\bibfnamefont
  {P.~Y.}\ \bibnamefont {Mai}}, \bibinfo {author} {\bibfnamefont {R.~C.~C.}\
  \bibnamefont {Leon}}, \bibinfo {author} {\bibfnamefont {C.~C.}\ \bibnamefont
  {Escott}}, \bibinfo {author} {\bibfnamefont {K.~M.}\ \bibnamefont {Itoh}},
  \bibinfo {author} {\bibfnamefont {N.~V.}\ \bibnamefont {Abrosimov}}, \bibinfo
  {author} {\bibfnamefont {H.-J.}\ \bibnamefont {Pohl}}, \bibinfo {author}
  {\bibfnamefont {M.~L.~W.}\ \bibnamefont {Thewalt}}, \bibinfo {author}
  {\bibfnamefont {F.~E.}\ \bibnamefont {Hudson}}, \bibinfo {author}
  {\bibfnamefont {A.}~\bibnamefont {Morello}}, \bibinfo {author} {\bibfnamefont
  {A.}~\bibnamefont {Laucht}}, \bibinfo {author} {\bibfnamefont {C.~H.}\
  \bibnamefont {Yang}}, \bibinfo {author} {\bibfnamefont {A.}~\bibnamefont
  {Saraiva}}, \ and\ \bibinfo {author} {\bibfnamefont {A.~S.}\ \bibnamefont
  {Dzurak}},\ }\href {\doibase https://doi.org/10.1038/s41565-022-01280-4}
  {\bibfield  {journal} {\bibinfo  {journal} {Nature Nanotechnology}\ }\textbf
  {\bibinfo {volume} {18}},\ \bibinfo {pages} {131–136} (\bibinfo {year}
  {2023})}\BibitemShut {NoStop}%
\bibitem [{\citenamefont {Harvey-Collard}\ \emph {et~al.}(2019)\citenamefont
  {Harvey-Collard}, \citenamefont {Jacobson}, \citenamefont {Bureau-Oxton},
  \citenamefont {Jock}, \citenamefont {Srinivasa}, \citenamefont {Mounce},
  \citenamefont {Ward}, \citenamefont {Anderson}, \citenamefont {Manginell},
  \citenamefont {Wendt}, \citenamefont {Pluym}, \citenamefont {Lilly},
  \citenamefont {Luhman}, \citenamefont {Pioro-Ladrière},\ and\ \citenamefont
  {Carroll}}]{harvey-collard_jacobson_bureau-oxton_jock_srinivasa_mounce_ward_anderson_manginell_wendt_etal_2019}%
  \BibitemOpen
  \bibfield  {author} {\bibinfo {author} {\bibfnamefont {P.}~\bibnamefont
  {Harvey-Collard}}, \bibinfo {author} {\bibfnamefont {N.~T.}\ \bibnamefont
  {Jacobson}}, \bibinfo {author} {\bibfnamefont {C.}~\bibnamefont
  {Bureau-Oxton}}, \bibinfo {author} {\bibfnamefont {R.~M.}\ \bibnamefont
  {Jock}}, \bibinfo {author} {\bibfnamefont {V.}~\bibnamefont {Srinivasa}},
  \bibinfo {author} {\bibfnamefont {A.~M.}\ \bibnamefont {Mounce}}, \bibinfo
  {author} {\bibfnamefont {D.~R.}\ \bibnamefont {Ward}}, \bibinfo {author}
  {\bibfnamefont {J.~M.}\ \bibnamefont {Anderson}}, \bibinfo {author}
  {\bibfnamefont {R.~P.}\ \bibnamefont {Manginell}}, \bibinfo {author}
  {\bibfnamefont {J.~R.}\ \bibnamefont {Wendt}}, \bibinfo {author}
  {\bibfnamefont {T.}~\bibnamefont {Pluym}}, \bibinfo {author} {\bibfnamefont
  {M.~P.}\ \bibnamefont {Lilly}}, \bibinfo {author} {\bibfnamefont {D.~R.}\
  \bibnamefont {Luhman}}, \bibinfo {author} {\bibfnamefont {M.}~\bibnamefont
  {Pioro-Ladrière}}, \ and\ \bibinfo {author} {\bibfnamefont {M.~S.}\
  \bibnamefont {Carroll}},\ }\href
  {https://journals.aps.org/prl/abstract/10.1103/PhysRevLett.122.217702}
  {\bibfield  {journal} {\bibinfo  {journal} {Physical Review Letters}\
  }\textbf {\bibinfo {volume} {122}} (\bibinfo {year} {2019})}\BibitemShut
  {NoStop}%
\bibitem [{\citenamefont {Tanttu}\ \emph {et~al.}(2019)\citenamefont {Tanttu},
  \citenamefont {Hensen}, \citenamefont {Chan}, \citenamefont {Yang},
  \citenamefont {Huang}, \citenamefont {Fogarty}, \citenamefont {Hudson},
  \citenamefont {Itoh}, \citenamefont {Culcer}, \citenamefont {Laucht},
  \citenamefont {Morello},\ and\ \citenamefont {Dzurak}}]{Tanttu_PRX19}%
  \BibitemOpen
  \bibfield  {author} {\bibinfo {author} {\bibfnamefont {T.}~\bibnamefont
  {Tanttu}}, \bibinfo {author} {\bibfnamefont {B.}~\bibnamefont {Hensen}},
  \bibinfo {author} {\bibfnamefont {K.~W.}\ \bibnamefont {Chan}}, \bibinfo
  {author} {\bibfnamefont {C.~H.}\ \bibnamefont {Yang}}, \bibinfo {author}
  {\bibfnamefont {W.~W.}\ \bibnamefont {Huang}}, \bibinfo {author}
  {\bibfnamefont {M.}~\bibnamefont {Fogarty}}, \bibinfo {author} {\bibfnamefont
  {F.}~\bibnamefont {Hudson}}, \bibinfo {author} {\bibfnamefont
  {K.}~\bibnamefont {Itoh}}, \bibinfo {author} {\bibfnamefont {D.}~\bibnamefont
  {Culcer}}, \bibinfo {author} {\bibfnamefont {A.}~\bibnamefont {Laucht}},
  \bibinfo {author} {\bibfnamefont {A.}~\bibnamefont {Morello}}, \ and\
  \bibinfo {author} {\bibfnamefont {A.}~\bibnamefont {Dzurak}},\ }\href
  {\doibase 10.1103/PhysRevX.9.021028} {\bibfield  {journal} {\bibinfo
  {journal} {Phys. Rev. X}\ }\textbf {\bibinfo {volume} {9}},\ \bibinfo {pages}
  {021028} (\bibinfo {year} {2019})}\BibitemShut {NoStop}%
\bibitem [{\citenamefont {Takeda}\ \emph {et~al.}(2022)\citenamefont {Takeda},
  \citenamefont {Noiri}, \citenamefont {Nakajima}, \citenamefont {Kobayashi},\
  and\ \citenamefont {Tarucha}}]{takeda_nat22}%
  \BibitemOpen
  \bibfield  {author} {\bibinfo {author} {\bibfnamefont {K.}~\bibnamefont
  {Takeda}}, \bibinfo {author} {\bibfnamefont {A.}~\bibnamefont {Noiri}},
  \bibinfo {author} {\bibfnamefont {T.}~\bibnamefont {Nakajima}}, \bibinfo
  {author} {\bibfnamefont {T.}~\bibnamefont {Kobayashi}}, \ and\ \bibinfo
  {author} {\bibfnamefont {S.}~\bibnamefont {Tarucha}},\ }\href@noop {}
  {\bibfield  {journal} {\bibinfo  {journal} {Nature}\ }\textbf {\bibinfo
  {volume} {608}},\ \bibinfo {pages} {682} (\bibinfo {year}
  {2022})}\BibitemShut {NoStop}%
\bibitem [{\citenamefont {Takeda}\ \emph {et~al.}(2016)\citenamefont {Takeda},
  \citenamefont {Kamioka}, \citenamefont {Otsuka}, \citenamefont {Yoneda},
  \citenamefont {Nakajima}, \citenamefont {Delbecq}, \citenamefont {Amaha},
  \citenamefont {Allison}, \citenamefont {Kodera}, \citenamefont {Oda},\ and\
  \citenamefont {{Seigo Tarucha}}}]{Takeda_SA16}%
  \BibitemOpen
  \bibfield  {author} {\bibinfo {author} {\bibfnamefont {K.}~\bibnamefont
  {Takeda}}, \bibinfo {author} {\bibfnamefont {J.}~\bibnamefont {Kamioka}},
  \bibinfo {author} {\bibfnamefont {T.}~\bibnamefont {Otsuka}}, \bibinfo
  {author} {\bibfnamefont {J.}~\bibnamefont {Yoneda}}, \bibinfo {author}
  {\bibfnamefont {T.}~\bibnamefont {Nakajima}}, \bibinfo {author}
  {\bibfnamefont {M.~R.}\ \bibnamefont {Delbecq}}, \bibinfo {author}
  {\bibfnamefont {S.}~\bibnamefont {Amaha}}, \bibinfo {author} {\bibfnamefont
  {G.}~\bibnamefont {Allison}}, \bibinfo {author} {\bibfnamefont
  {T.}~\bibnamefont {Kodera}}, \bibinfo {author} {\bibfnamefont
  {S.}~\bibnamefont {Oda}}, \ and\ \bibinfo {author} {\bibnamefont {{Seigo
  Tarucha}}},\ }\href {\doibase 10.1126/sciadv.1600694} {\bibfield  {journal}
  {\bibinfo  {journal} {Science Advances}\ }\textbf {\bibinfo {volume} {2}},\
  \bibinfo {pages} {e1600694} (\bibinfo {year} {2016})}\BibitemShut {NoStop}%
\bibitem [{\citenamefont {Brunner}\ \emph {et~al.}(2011)\citenamefont
  {Brunner}, \citenamefont {Shin}, \citenamefont {Obata}, \citenamefont
  {Pioro-Ladri{\`e}re}, \citenamefont {Kubo}, \citenamefont {Yoshida},
  \citenamefont {Taniyama}, \citenamefont {Tokura},\ and\ \citenamefont
  {Tarucha}}]{Brunner_PRL11}%
  \BibitemOpen
  \bibfield  {author} {\bibinfo {author} {\bibfnamefont {R.}~\bibnamefont
  {Brunner}}, \bibinfo {author} {\bibfnamefont {Y.-S.}\ \bibnamefont {Shin}},
  \bibinfo {author} {\bibfnamefont {T.}~\bibnamefont {Obata}}, \bibinfo
  {author} {\bibfnamefont {M.}~\bibnamefont {Pioro-Ladri{\`e}re}}, \bibinfo
  {author} {\bibfnamefont {T.}~\bibnamefont {Kubo}}, \bibinfo {author}
  {\bibfnamefont {K.}~\bibnamefont {Yoshida}}, \bibinfo {author} {\bibfnamefont
  {T.}~\bibnamefont {Taniyama}}, \bibinfo {author} {\bibfnamefont
  {Y.}~\bibnamefont {Tokura}}, \ and\ \bibinfo {author} {\bibfnamefont
  {S.}~\bibnamefont {Tarucha}},\ }\href {\doibase
  10.1103/PhysRevLett.107.146801} {\bibfield  {journal} {\bibinfo  {journal}
  {Phys.~Rev.~Lett.}\ }\textbf {\bibinfo {volume} {107}},\ \bibinfo {pages}
  {146801} (\bibinfo {year} {2011})}\BibitemShut {NoStop}%
\bibitem [{\citenamefont {Connors}\ \emph {et~al.}(2019)\citenamefont
  {Connors}, \citenamefont {Nelson}, \citenamefont {Qiao}, \citenamefont
  {Edge},\ and\ \citenamefont {Nichol}}]{Connors_PRB19}%
  \BibitemOpen
  \bibfield  {author} {\bibinfo {author} {\bibfnamefont {E.~J.}\ \bibnamefont
  {Connors}}, \bibinfo {author} {\bibfnamefont {J.}~\bibnamefont {Nelson}},
  \bibinfo {author} {\bibfnamefont {H.}~\bibnamefont {Qiao}}, \bibinfo {author}
  {\bibfnamefont {L.~F.}\ \bibnamefont {Edge}}, \ and\ \bibinfo {author}
  {\bibfnamefont {J.~M.}\ \bibnamefont {Nichol}},\ }\href {\doibase
  10.1103/PhysRevB.100.165305} {\bibfield  {journal} {\bibinfo  {journal}
  {Phys.~Rev.~B}\ }\textbf {\bibinfo {volume} {100}},\ \bibinfo {pages}
  {165305} (\bibinfo {year} {2019})}\BibitemShut {NoStop}%
\bibitem [{\citenamefont {Yoneda}\ \emph {et~al.}(2022)\citenamefont {Yoneda},
  \citenamefont {{Rojas-Arias}}, \citenamefont {Stano}, \citenamefont {Takeda},
  \citenamefont {Noiri}, \citenamefont {Nakajima}, \citenamefont {Loss},\ and\
  \citenamefont {Tarucha}}]{Yoneda22}%
  \BibitemOpen
  \bibfield  {author} {\bibinfo {author} {\bibfnamefont {J.}~\bibnamefont
  {Yoneda}}, \bibinfo {author} {\bibfnamefont {J.~S.}\ \bibnamefont
  {{Rojas-Arias}}}, \bibinfo {author} {\bibfnamefont {P.}~\bibnamefont
  {Stano}}, \bibinfo {author} {\bibfnamefont {K.}~\bibnamefont {Takeda}},
  \bibinfo {author} {\bibfnamefont {A.}~\bibnamefont {Noiri}}, \bibinfo
  {author} {\bibfnamefont {T.}~\bibnamefont {Nakajima}}, \bibinfo {author}
  {\bibfnamefont {D.}~\bibnamefont {Loss}}, \ and\ \bibinfo {author}
  {\bibfnamefont {S.}~\bibnamefont {Tarucha}},\ }\href
  {https://arxiv.org/abs/2208.14150} {\bibfield  {journal} {\bibinfo  {journal}
  {arXiv:2208.14150}\ } (\bibinfo {year} {2022})}\BibitemShut {NoStop}%
\bibitem [{\citenamefont {Petersson}\ \emph {et~al.}(2010)\citenamefont
  {Petersson}, \citenamefont {Petta}, \citenamefont {Lu},\ and\ \citenamefont
  {Gossard}}]{petersson_petta_lu_gossard_2010}%
  \BibitemOpen
  \bibfield  {author} {\bibinfo {author} {\bibfnamefont {K.~D.}\ \bibnamefont
  {Petersson}}, \bibinfo {author} {\bibfnamefont {J.~R.}\ \bibnamefont
  {Petta}}, \bibinfo {author} {\bibfnamefont {H.}~\bibnamefont {Lu}}, \ and\
  \bibinfo {author} {\bibfnamefont {A.~C.}\ \bibnamefont {Gossard}},\ }\href
  {\doibase https://doi.org/10.1103/physrevlett.105.246804} {\bibfield
  {journal} {\bibinfo  {journal} {Phys.~Rev.~Lett.}\ }\textbf {\bibinfo
  {volume} {105}} (\bibinfo {year} {2010}),\
  https://doi.org/10.1103/physrevlett.105.246804}\BibitemShut {NoStop}%
\bibitem [{\citenamefont {Mi}\ \emph {et~al.}(2018)\citenamefont {Mi},
  \citenamefont {Kohler},\ and\ \citenamefont {Petta}}]{Mi_PRB18}%
  \BibitemOpen
  \bibfield  {author} {\bibinfo {author} {\bibfnamefont {X.}~\bibnamefont
  {Mi}}, \bibinfo {author} {\bibfnamefont {S.}~\bibnamefont {Kohler}}, \ and\
  \bibinfo {author} {\bibfnamefont {J.~R.}\ \bibnamefont {Petta}},\ }\href
  {\doibase 10.1103/PhysRevB.98.161404} {\bibfield  {journal} {\bibinfo
  {journal} {Phys. Rev. B}\ }\textbf {\bibinfo {volume} {98}},\ \bibinfo
  {pages} {161404} (\bibinfo {year} {2018})}\BibitemShut {NoStop}%
\bibitem [{\citenamefont {Shehata}\ \emph {et~al.}(2022)\citenamefont
  {Shehata}, \citenamefont {Simion}, \citenamefont {Li}, \citenamefont
  {Mohiyaddin}, \citenamefont {Wan}, \citenamefont {Mongillo}, \citenamefont
  {Govoreanu}, \citenamefont {Radu}, \citenamefont {De~Greve},\ and\
  \citenamefont {Van~Dorpe}}]{Shehata22}%
  \BibitemOpen
  \bibfield  {author} {\bibinfo {author} {\bibfnamefont {M.~M. E.~K.}\
  \bibnamefont {Shehata}}, \bibinfo {author} {\bibfnamefont {G.}~\bibnamefont
  {Simion}}, \bibinfo {author} {\bibfnamefont {R.}~\bibnamefont {Li}}, \bibinfo
  {author} {\bibfnamefont {F.~A.}\ \bibnamefont {Mohiyaddin}}, \bibinfo
  {author} {\bibfnamefont {D.}~\bibnamefont {Wan}}, \bibinfo {author}
  {\bibfnamefont {M.}~\bibnamefont {Mongillo}}, \bibinfo {author}
  {\bibfnamefont {B.}~\bibnamefont {Govoreanu}}, \bibinfo {author}
  {\bibfnamefont {I.}~\bibnamefont {Radu}}, \bibinfo {author} {\bibfnamefont
  {K.}~\bibnamefont {De~Greve}}, \ and\ \bibinfo {author} {\bibfnamefont
  {P.}~\bibnamefont {Van~Dorpe}},\ }\href {https://arxiv.org/abs/2210.04539}
  {\bibfield  {journal} {\bibinfo  {journal} {arXiv:2210.04539}\ } (\bibinfo
  {year} {2022})}\BibitemShut {NoStop}%
\bibitem [{\citenamefont {Shalak}\ \emph {et~al.}(2022)\citenamefont {Shalak},
  \citenamefont {Delerue},\ and\ \citenamefont {Niquet}}]{Shalak22}%
  \BibitemOpen
  \bibfield  {author} {\bibinfo {author} {\bibfnamefont {B.}~\bibnamefont
  {Shalak}}, \bibinfo {author} {\bibfnamefont {C.}~\bibnamefont {Delerue}}, \
  and\ \bibinfo {author} {\bibfnamefont {Y.-M.}\ \bibnamefont {Niquet}},\
  }\href {http://arxiv.org/abs/2210.10476} {\bibfield  {journal} {\bibinfo
  {journal} {arXiv:2210.10476}\ } (\bibinfo {year} {2022})}\BibitemShut
  {NoStop}%
\bibitem [{\citenamefont {Seidler}\ \emph {et~al.}(2022)\citenamefont
  {Seidler}, \citenamefont {Struck}, \citenamefont {Xue}, \citenamefont
  {Focke}, \citenamefont {Trellenkamp}, \citenamefont {Bluhm},\ and\
  \citenamefont {Schreiber}}]{Seidler_NPJQI22}%
  \BibitemOpen
  \bibfield  {author} {\bibinfo {author} {\bibfnamefont {I.}~\bibnamefont
  {Seidler}}, \bibinfo {author} {\bibfnamefont {T.}~\bibnamefont {Struck}},
  \bibinfo {author} {\bibfnamefont {R.}~\bibnamefont {Xue}}, \bibinfo {author}
  {\bibfnamefont {N.}~\bibnamefont {Focke}}, \bibinfo {author} {\bibfnamefont
  {S.}~\bibnamefont {Trellenkamp}}, \bibinfo {author} {\bibfnamefont
  {H.}~\bibnamefont {Bluhm}}, \ and\ \bibinfo {author} {\bibfnamefont {L.~R.}\
  \bibnamefont {Schreiber}},\ }\href {\doibase 10.1038/s41534-022-00615-2}
  {\bibfield  {journal} {\bibinfo  {journal} {npj Quantum Information}\
  }\textbf {\bibinfo {volume} {8}},\ \bibinfo {pages} {100} (\bibinfo {year}
  {2022})}\BibitemShut {NoStop}%
\bibitem [{\citenamefont {Langrock}\ \emph {et~al.}(2022)\citenamefont
  {Langrock}, \citenamefont {Krzywda}, \citenamefont {Focke}, \citenamefont
  {Seidler}, \citenamefont {Schreiber},\ and\ \citenamefont
  {Cywi{\'n}ski}}]{Langrock22}%
  \BibitemOpen
  \bibfield  {author} {\bibinfo {author} {\bibfnamefont {V.}~\bibnamefont
  {Langrock}}, \bibinfo {author} {\bibfnamefont {J.~A.}\ \bibnamefont
  {Krzywda}}, \bibinfo {author} {\bibfnamefont {N.}~\bibnamefont {Focke}},
  \bibinfo {author} {\bibfnamefont {I.}~\bibnamefont {Seidler}}, \bibinfo
  {author} {\bibfnamefont {L.~R.}\ \bibnamefont {Schreiber}}, \ and\ \bibinfo
  {author} {\bibfnamefont {{\L}.}~\bibnamefont {Cywi{\'n}ski}},\ }\href
  {https://arxiv.org/abs/2202.11793} {\bibfield  {journal} {\bibinfo  {journal}
  {arXiv:2202.11793}\ } (\bibinfo {year} {2022})}\BibitemShut {NoStop}%
\bibitem [{\citenamefont {Klos}\ \emph {et~al.}(2018)\citenamefont {Klos},
  \citenamefont {Hassler}, \citenamefont {Cerfontaine}, \citenamefont {Bluhm},\
  and\ \citenamefont {Schreiber}}]{klosCalculationTunnelCouplings2018}%
  \BibitemOpen
  \bibfield  {author} {\bibinfo {author} {\bibfnamefont {J.}~\bibnamefont
  {Klos}}, \bibinfo {author} {\bibfnamefont {F.}~\bibnamefont {Hassler}},
  \bibinfo {author} {\bibfnamefont {P.}~\bibnamefont {Cerfontaine}}, \bibinfo
  {author} {\bibfnamefont {H.}~\bibnamefont {Bluhm}}, \ and\ \bibinfo {author}
  {\bibfnamefont {L.~R.}\ \bibnamefont {Schreiber}},\ }\href {\doibase
  10.1103/PhysRevB.98.155320} {\bibfield  {journal} {\bibinfo  {journal}
  {Physical Review B}\ }\textbf {\bibinfo {volume} {98}},\ \bibinfo {pages}
  {155320} (\bibinfo {year} {2018})}\BibitemShut {NoStop}%
\bibitem [{\citenamefont {Klos}\ \emph {et~al.}(2019)\citenamefont {Klos},
  \citenamefont {Sun}, \citenamefont {Beyer}, \citenamefont {Kindel},
  \citenamefont {Hellmich}, \citenamefont {Knoch},\ and\ \citenamefont
  {Schreiber}}]{klosSpinQubitsConfined2019}%
  \BibitemOpen
  \bibfield  {author} {\bibinfo {author} {\bibfnamefont {J.}~\bibnamefont
  {Klos}}, \bibinfo {author} {\bibfnamefont {B.}~\bibnamefont {Sun}}, \bibinfo
  {author} {\bibfnamefont {J.}~\bibnamefont {Beyer}}, \bibinfo {author}
  {\bibfnamefont {S.}~\bibnamefont {Kindel}}, \bibinfo {author} {\bibfnamefont
  {L.}~\bibnamefont {Hellmich}}, \bibinfo {author} {\bibfnamefont
  {J.}~\bibnamefont {Knoch}}, \ and\ \bibinfo {author} {\bibfnamefont {L.~R.}\
  \bibnamefont {Schreiber}},\ }\href {\doibase 10.3390/app9183823} {\bibfield
  {journal} {\bibinfo  {journal} {Applied Sciences}\ }\textbf {\bibinfo
  {volume} {9}},\ \bibinfo {pages} {3823} (\bibinfo {year} {2019})}\BibitemShut
  {NoStop}%
\bibitem [{\citenamefont {Spruijtenburg}\ \emph {et~al.}(2016)\citenamefont
  {Spruijtenburg}, \citenamefont {Amitonov}, \citenamefont {Mueller},
  \citenamefont {{van der Wiel}},\ and\ \citenamefont
  {Zwanenburg}}]{spruijtenburgPassivationCharacterizationCharge2016}%
  \BibitemOpen
  \bibfield  {author} {\bibinfo {author} {\bibfnamefont {P.~C.}\ \bibnamefont
  {Spruijtenburg}}, \bibinfo {author} {\bibfnamefont {S.~V.}\ \bibnamefont
  {Amitonov}}, \bibinfo {author} {\bibfnamefont {F.}~\bibnamefont {Mueller}},
  \bibinfo {author} {\bibfnamefont {W.~G.}\ \bibnamefont {{van der Wiel}}}, \
  and\ \bibinfo {author} {\bibfnamefont {F.~A.}\ \bibnamefont {Zwanenburg}},\
  }\href {\doibase 10.1038/srep38127} {\bibfield  {journal} {\bibinfo
  {journal} {Scientific Reports}\ }\textbf {\bibinfo {volume} {6}},\ \bibinfo
  {pages} {38127} (\bibinfo {year} {2016})}\BibitemShut {NoStop}%
\bibitem [{\citenamefont {Thoan}\ \emph {et~al.}(2011)\citenamefont {Thoan},
  \citenamefont {Keunen}, \citenamefont {Afanas'ev},\ and\ \citenamefont
  {Stesmans}}]{thoanInterfaceStateEnergy2011}%
  \BibitemOpen
  \bibfield  {author} {\bibinfo {author} {\bibfnamefont {N.~H.}\ \bibnamefont
  {Thoan}}, \bibinfo {author} {\bibfnamefont {K.}~\bibnamefont {Keunen}},
  \bibinfo {author} {\bibfnamefont {V.~V.}\ \bibnamefont {Afanas'ev}}, \ and\
  \bibinfo {author} {\bibfnamefont {A.}~\bibnamefont {Stesmans}},\ }\href
  {\doibase 10.1063/1.3527909} {\bibfield  {journal} {\bibinfo  {journal}
  {Journal of Applied Physics}\ }\textbf {\bibinfo {volume} {109}},\ \bibinfo
  {pages} {013710} (\bibinfo {year} {2011})}\BibitemShut {NoStop}%
\bibitem [{\citenamefont {Campbell}\ \emph {et~al.}(2005)\citenamefont
  {Campbell}, \citenamefont {Lenahan}, \citenamefont {Krishnan},\ and\
  \citenamefont {Krishnan}}]{Campbell05}%
  \BibitemOpen
  \bibfield  {author} {\bibinfo {author} {\bibfnamefont {J.~P.}\ \bibnamefont
  {Campbell}}, \bibinfo {author} {\bibfnamefont {P.~M.}\ \bibnamefont
  {Lenahan}}, \bibinfo {author} {\bibfnamefont {A.~T.}\ \bibnamefont
  {Krishnan}}, \ and\ \bibinfo {author} {\bibfnamefont {S.}~\bibnamefont
  {Krishnan}},\ }\href {\doibase 10.1063/1.2131197} {\bibfield  {journal}
  {\bibinfo  {journal} {Appl. Phys. Lett.}\ }\textbf {\bibinfo {volume} {87}},\
  \bibinfo {pages} {204106} (\bibinfo {year} {2005})}\BibitemShut {NoStop}%
\bibitem [{\citenamefont {Machlup}(1954)}]{Machlup_JAP54}%
  \BibitemOpen
  \bibfield  {author} {\bibinfo {author} {\bibfnamefont {S.}~\bibnamefont
  {Machlup}},\ }\href {\doibase 10.1063/1.1721637} {\bibfield  {journal}
  {\bibinfo  {journal} {Journal of Applied Physics}\ }\textbf {\bibinfo
  {volume} {25}},\ \bibinfo {pages} {341} (\bibinfo {year} {1954})}\BibitemShut
  {NoStop}%
\bibitem [{\citenamefont {Constantin}\ \emph {et~al.}(2009)\citenamefont
  {Constantin}, \citenamefont {Yu},\ and\ \citenamefont
  {Martinis}}]{Constantin_PRB09}%
  \BibitemOpen
  \bibfield  {author} {\bibinfo {author} {\bibfnamefont {M.}~\bibnamefont
  {Constantin}}, \bibinfo {author} {\bibfnamefont {C.~C.}\ \bibnamefont {Yu}},
  \ and\ \bibinfo {author} {\bibfnamefont {J.~M.}\ \bibnamefont {Martinis}},\
  }\href {\doibase 10.1103/PhysRevB.79.094520} {\bibfield  {journal} {\bibinfo
  {journal} {Phys. Rev. B}\ }\textbf {\bibinfo {volume} {79}},\ \bibinfo
  {pages} {094520} (\bibinfo {year} {2009})}\BibitemShut {NoStop}%
\bibitem [{\citenamefont {Schriefl}\ \emph {et~al.}(2006)\citenamefont
  {Schriefl}, \citenamefont {Makhlin}, \citenamefont {Shnirman},\ and\
  \citenamefont {Sch{\"o}n}}]{schriefl2006decoherence}%
  \BibitemOpen
  \bibfield  {author} {\bibinfo {author} {\bibfnamefont {J.}~\bibnamefont
  {Schriefl}}, \bibinfo {author} {\bibfnamefont {Y.}~\bibnamefont {Makhlin}},
  \bibinfo {author} {\bibfnamefont {A.}~\bibnamefont {Shnirman}}, \ and\
  \bibinfo {author} {\bibfnamefont {G.}~\bibnamefont {Sch{\"o}n}},\ }\href@noop
  {} {\bibfield  {journal} {\bibinfo  {journal} {New Journal of Physics}\
  }\textbf {\bibinfo {volume} {8}},\ \bibinfo {pages} {1} (\bibinfo {year}
  {2006})}\BibitemShut {NoStop}%
\bibitem [{\citenamefont {You}\ \emph {et~al.}(2021)\citenamefont {You},
  \citenamefont {Clerk},\ and\ \citenamefont {Koch}}]{YouPRR21}%
  \BibitemOpen
  \bibfield  {author} {\bibinfo {author} {\bibfnamefont {X.}~\bibnamefont
  {You}}, \bibinfo {author} {\bibfnamefont {A.~A.}\ \bibnamefont {Clerk}}, \
  and\ \bibinfo {author} {\bibfnamefont {J.}~\bibnamefont {Koch}},\ }\href
  {\doibase 10.1103/PhysRevResearch.3.013045} {\bibfield  {journal} {\bibinfo
  {journal} {Phys. Rev. Res.}\ }\textbf {\bibinfo {volume} {3}},\ \bibinfo
  {pages} {013045} (\bibinfo {year} {2021})}\BibitemShut {NoStop}%
\bibitem [{\citenamefont {Shnirman}\ \emph {et~al.}(2005)\citenamefont
  {Shnirman}, \citenamefont {Sch\"on}, \citenamefont {Martin},\ and\
  \citenamefont {Makhlin}}]{Shnirman_PRL05}%
  \BibitemOpen
  \bibfield  {author} {\bibinfo {author} {\bibfnamefont {A.}~\bibnamefont
  {Shnirman}}, \bibinfo {author} {\bibfnamefont {G.}~\bibnamefont {Sch\"on}},
  \bibinfo {author} {\bibfnamefont {I.}~\bibnamefont {Martin}}, \ and\ \bibinfo
  {author} {\bibfnamefont {Y.}~\bibnamefont {Makhlin}},\ }\href {\doibase
  10.1103/PhysRevLett.94.127002} {\bibfield  {journal} {\bibinfo  {journal}
  {Phys. Rev. Lett.}\ }\textbf {\bibinfo {volume} {94}},\ \bibinfo {pages}
  {127002} (\bibinfo {year} {2005})}\BibitemShut {NoStop}%
\bibitem [{\citenamefont {Müller}\ \emph {et~al.}(2019)\citenamefont
  {Müller}, \citenamefont {Cole},\ and\ \citenamefont
  {Lisenfeld}}]{Muller_2019}%
  \BibitemOpen
  \bibfield  {author} {\bibinfo {author} {\bibfnamefont {C.}~\bibnamefont
  {Müller}}, \bibinfo {author} {\bibfnamefont {J.~H.}\ \bibnamefont {Cole}}, \
  and\ \bibinfo {author} {\bibfnamefont {J.}~\bibnamefont {Lisenfeld}},\ }\href
  {\doibase 10.1088/1361-6633/ab3a7e} {\bibfield  {journal} {\bibinfo
  {journal} {Reports on Progress in Physics}\ }\textbf {\bibinfo {volume}
  {82}},\ \bibinfo {pages} {124501} (\bibinfo {year} {2019})}\BibitemShut
  {NoStop}%
\bibitem [{\citenamefont {Beaudoin}\ and\ \citenamefont
  {Coish}(2015)}]{Beaudoin_PRB15}%
  \BibitemOpen
  \bibfield  {author} {\bibinfo {author} {\bibfnamefont {F.}~\bibnamefont
  {Beaudoin}}\ and\ \bibinfo {author} {\bibfnamefont {W.~A.}\ \bibnamefont
  {Coish}},\ }\href {\doibase 10.1103/PhysRevB.91.165432} {\bibfield  {journal}
  {\bibinfo  {journal} {Phys. Rev. B}\ }\textbf {\bibinfo {volume} {91}},\
  \bibinfo {pages} {165432} (\bibinfo {year} {2015})}\BibitemShut {NoStop}%
\bibitem [{\citenamefont {Ahn}\ \emph {et~al.}(2021)\citenamefont {Ahn},
  \citenamefont {Das~Sarma},\ and\ \citenamefont {Kestner}}]{Ahn_PRB21}%
  \BibitemOpen
  \bibfield  {author} {\bibinfo {author} {\bibfnamefont {S.}~\bibnamefont
  {Ahn}}, \bibinfo {author} {\bibfnamefont {S.}~\bibnamefont {Das~Sarma}}, \
  and\ \bibinfo {author} {\bibfnamefont {J.~P.}\ \bibnamefont {Kestner}},\
  }\href {\doibase 10.1103/PhysRevB.103.L041304} {\bibfield  {journal}
  {\bibinfo  {journal} {Phys. Rev. B}\ }\textbf {\bibinfo {volume} {103}},\
  \bibinfo {pages} {L041304} (\bibinfo {year} {2021})}\BibitemShut {NoStop}%
\bibitem [{\citenamefont {Kranz}\ \emph {et~al.}(2020)\citenamefont {Kranz},
  \citenamefont {Gorman}, \citenamefont {Thorgrimsson}, \citenamefont {He},
  \citenamefont {Keith}, \citenamefont {Keizer},\ and\ \citenamefont
  {Simmons}}]{KranzAM20}%
  \BibitemOpen
  \bibfield  {author} {\bibinfo {author} {\bibfnamefont {L.}~\bibnamefont
  {Kranz}}, \bibinfo {author} {\bibfnamefont {S.~K.}\ \bibnamefont {Gorman}},
  \bibinfo {author} {\bibfnamefont {B.}~\bibnamefont {Thorgrimsson}}, \bibinfo
  {author} {\bibfnamefont {Y.}~\bibnamefont {He}}, \bibinfo {author}
  {\bibfnamefont {D.}~\bibnamefont {Keith}}, \bibinfo {author} {\bibfnamefont
  {J.~G.}\ \bibnamefont {Keizer}}, \ and\ \bibinfo {author} {\bibfnamefont
  {M.~Y.}\ \bibnamefont {Simmons}},\ }\href {\doibase
  https://doi.org/10.1002/adma.202003361} {\bibfield  {journal} {\bibinfo
  {journal} {Advanced Materials}\ }\textbf {\bibinfo {volume} {32}},\ \bibinfo
  {pages} {2003361} (\bibinfo {year} {2020})}\BibitemShut {NoStop}%
\bibitem [{\citenamefont {Mills}\ \emph {et~al.}(2022)\citenamefont {Mills},
  \citenamefont {Guinn}, \citenamefont {Gullans}, \citenamefont {Sigillito},
  \citenamefont {Feldman}, \citenamefont {Nielsen},\ and\ \citenamefont
  {Petta1}}]{Mills_SA22}%
  \BibitemOpen
  \bibfield  {author} {\bibinfo {author} {\bibfnamefont {A.~R.}\ \bibnamefont
  {Mills}}, \bibinfo {author} {\bibfnamefont {C.~R.}\ \bibnamefont {Guinn}},
  \bibinfo {author} {\bibfnamefont {M.~J.}\ \bibnamefont {Gullans}}, \bibinfo
  {author} {\bibfnamefont {A.~J.}\ \bibnamefont {Sigillito}}, \bibinfo {author}
  {\bibfnamefont {M.~M.}\ \bibnamefont {Feldman}}, \bibinfo {author}
  {\bibfnamefont {E.}~\bibnamefont {Nielsen}}, \ and\ \bibinfo {author}
  {\bibfnamefont {J.~R.}\ \bibnamefont {Petta1}},\ }\href {\doibase
  10.1126/sciadv.abn5130} {\bibfield  {journal} {\bibinfo  {journal}
  {Sci.~Adv.}\ }\textbf {\bibinfo {volume} {8}},\ \bibinfo {pages} {eabn5130}
  (\bibinfo {year} {2022})}\BibitemShut {NoStop}%
\bibitem [{\citenamefont {Zajac}\ \emph {et~al.}(2015)\citenamefont {Zajac},
  \citenamefont {Hazard}, \citenamefont {Mi}, \citenamefont {Wang},\ and\
  \citenamefont {Petta}}]{zajac_hazard_mi_wang_petta_2015}%
  \BibitemOpen
  \bibfield  {author} {\bibinfo {author} {\bibfnamefont {D.~M.}\ \bibnamefont
  {Zajac}}, \bibinfo {author} {\bibfnamefont {T.~M.}\ \bibnamefont {Hazard}},
  \bibinfo {author} {\bibfnamefont {X.}~\bibnamefont {Mi}}, \bibinfo {author}
  {\bibfnamefont {K.}~\bibnamefont {Wang}}, \ and\ \bibinfo {author}
  {\bibfnamefont {J.~R.}\ \bibnamefont {Petta}},\ }\href {\doibase
  https://doi.org/10.1063/1.4922249} {\bibfield  {journal} {\bibinfo  {journal}
  {Applied Physics Letters}\ }\textbf {\bibinfo {volume} {106}},\ \bibinfo
  {pages} {223507} (\bibinfo {year} {2015})}\BibitemShut {NoStop}%
\bibitem [{\citenamefont {Borselli}\ \emph {et~al.}(2015)\citenamefont
  {Borselli}, \citenamefont {Eng}, \citenamefont {Ross}, \citenamefont
  {Hazard}, \citenamefont {Holabird}, \citenamefont {Huang}, \citenamefont
  {Kiselev}, \citenamefont {Deelman}, \citenamefont {Warren}, \citenamefont
  {Milosavljevic}, \citenamefont {Schmitz}, \citenamefont {Sokolich},
  \citenamefont {Gyure},\ and\ \citenamefont {Hunter}}]{Borselli_2015}%
  \BibitemOpen
  \bibfield  {author} {\bibinfo {author} {\bibfnamefont {M.~G.}\ \bibnamefont
  {Borselli}}, \bibinfo {author} {\bibfnamefont {K.}~\bibnamefont {Eng}},
  \bibinfo {author} {\bibfnamefont {R.~S.}\ \bibnamefont {Ross}}, \bibinfo
  {author} {\bibfnamefont {T.~M.}\ \bibnamefont {Hazard}}, \bibinfo {author}
  {\bibfnamefont {K.~S.}\ \bibnamefont {Holabird}}, \bibinfo {author}
  {\bibfnamefont {B.}~\bibnamefont {Huang}}, \bibinfo {author} {\bibfnamefont
  {A.~A.}\ \bibnamefont {Kiselev}}, \bibinfo {author} {\bibfnamefont {P.~W.}\
  \bibnamefont {Deelman}}, \bibinfo {author} {\bibfnamefont {L.~D.}\
  \bibnamefont {Warren}}, \bibinfo {author} {\bibfnamefont {I.}~\bibnamefont
  {Milosavljevic}}, \bibinfo {author} {\bibfnamefont {A.~E.}\ \bibnamefont
  {Schmitz}}, \bibinfo {author} {\bibfnamefont {M.}~\bibnamefont {Sokolich}},
  \bibinfo {author} {\bibfnamefont {M.~F.}\ \bibnamefont {Gyure}}, \ and\
  \bibinfo {author} {\bibfnamefont {A.~T.}\ \bibnamefont {Hunter}},\ }\href
  {\doibase 10.1088/0957-4484/26/37/375202} {\bibfield  {journal} {\bibinfo
  {journal} {Nanotechnology}\ }\textbf {\bibinfo {volume} {26}},\ \bibinfo
  {pages} {375202} (\bibinfo {year} {2015})}\BibitemShut {NoStop}%
\bibitem [{\citenamefont {Paquelet~Wuetz}\ \emph {et~al.}(2022)\citenamefont
  {Paquelet~Wuetz}, \citenamefont {Losert}, \citenamefont {Koelling},
  \citenamefont {Stehouwer}, \citenamefont {Zwerver}, \citenamefont {Philips},
  \citenamefont {Madzik}, \citenamefont {Xue}, \citenamefont {Zheng},
  \citenamefont {Lodari}, \citenamefont {Amitonov}, \citenamefont
  {Samkharadze}, \citenamefont {Sammak}, \citenamefont {Vandersypen},
  \citenamefont {Rahman}, \citenamefont {Coppersmith}, \citenamefont
  {Moutanabbir}, \citenamefont {Friesen},\ and\ \citenamefont
  {Scappucci}}]{Wuetz_NC22}%
  \BibitemOpen
  \bibfield  {author} {\bibinfo {author} {\bibfnamefont {B.}~\bibnamefont
  {Paquelet~Wuetz}}, \bibinfo {author} {\bibfnamefont {M.~P.}\ \bibnamefont
  {Losert}}, \bibinfo {author} {\bibfnamefont {S.}~\bibnamefont {Koelling}},
  \bibinfo {author} {\bibfnamefont {L.~E.~A.}\ \bibnamefont {Stehouwer}},
  \bibinfo {author} {\bibfnamefont {A.-M.~J.}\ \bibnamefont {Zwerver}},
  \bibinfo {author} {\bibfnamefont {S.~G.~J.}\ \bibnamefont {Philips}},
  \bibinfo {author} {\bibfnamefont {M.~T.}\ \bibnamefont {Madzik}}, \bibinfo
  {author} {\bibfnamefont {X.}~\bibnamefont {Xue}}, \bibinfo {author}
  {\bibfnamefont {G.}~\bibnamefont {Zheng}}, \bibinfo {author} {\bibfnamefont
  {M.}~\bibnamefont {Lodari}}, \bibinfo {author} {\bibfnamefont {S.~V.}\
  \bibnamefont {Amitonov}}, \bibinfo {author} {\bibfnamefont {N.}~\bibnamefont
  {Samkharadze}}, \bibinfo {author} {\bibfnamefont {A.}~\bibnamefont {Sammak}},
  \bibinfo {author} {\bibfnamefont {L.~M.~K.}\ \bibnamefont {Vandersypen}},
  \bibinfo {author} {\bibfnamefont {R.}~\bibnamefont {Rahman}}, \bibinfo
  {author} {\bibfnamefont {S.~N.}\ \bibnamefont {Coppersmith}}, \bibinfo
  {author} {\bibfnamefont {O.}~\bibnamefont {Moutanabbir}}, \bibinfo {author}
  {\bibfnamefont {M.}~\bibnamefont {Friesen}}, \ and\ \bibinfo {author}
  {\bibfnamefont {G.}~\bibnamefont {Scappucci}},\ }\href {\doibase
  10.1038/s41467-022-35458-0} {\bibfield  {journal} {\bibinfo  {journal} {Nat.
  Comm.}\ }\textbf {\bibinfo {volume} {13}},\ \bibinfo {pages} {7730} (\bibinfo
  {year} {2022})}\BibitemShut {NoStop}%
\bibitem [{\citenamefont {Holman}\ \emph {et~al.}(2021)\citenamefont {Holman},
  \citenamefont {Rosenberg}, \citenamefont {Yost}, \citenamefont {Yoder},
  \citenamefont {Das}, \citenamefont {Oliver}, \citenamefont {McDermott},\ and\
  \citenamefont
  {Eriksson}}]{holman_rosenberg_yost_yoder_das_oliver_mcdermott_eriksson_2021}%
  \BibitemOpen
  \bibfield  {author} {\bibinfo {author} {\bibfnamefont {N.}~\bibnamefont
  {Holman}}, \bibinfo {author} {\bibfnamefont {D.}~\bibnamefont {Rosenberg}},
  \bibinfo {author} {\bibfnamefont {D.}~\bibnamefont {Yost}}, \bibinfo {author}
  {\bibfnamefont {J.~L.}\ \bibnamefont {Yoder}}, \bibinfo {author}
  {\bibfnamefont {R.}~\bibnamefont {Das}}, \bibinfo {author} {\bibfnamefont
  {W.~D.}\ \bibnamefont {Oliver}}, \bibinfo {author} {\bibfnamefont
  {R.}~\bibnamefont {McDermott}}, \ and\ \bibinfo {author} {\bibfnamefont
  {M.~A.}\ \bibnamefont {Eriksson}},\ }\href {\doibase
  10.1038/s41534-021-00469-0} {\bibfield  {journal} {\bibinfo  {journal} {npj
  Quantum Information}\ }\textbf {\bibinfo {volume} {7}},\ \bibinfo {pages}
  {137} (\bibinfo {year} {2021})}\BibitemShut {NoStop}%
\bibitem [{\citenamefont {Zimmermann}\ and\ \citenamefont
  {Weber}(1981)}]{ZimmermannPRL1981}%
  \BibitemOpen
  \bibfield  {author} {\bibinfo {author} {\bibfnamefont {J.}~\bibnamefont
  {Zimmermann}}\ and\ \bibinfo {author} {\bibfnamefont {G.}~\bibnamefont
  {Weber}},\ }\href {\doibase 10.1103/PhysRevLett.46.661} {\bibfield  {journal}
  {\bibinfo  {journal} {Physical Review Letters}\ }\textbf {\bibinfo {volume}
  {46}},\ \bibinfo {pages} {661} (\bibinfo {year} {1981})}\BibitemShut
  {NoStop}%
\bibitem [{\citenamefont {Shi}\ \emph {et~al.}(2013)\citenamefont {Shi},
  \citenamefont {Simmons}, \citenamefont {Ward}, \citenamefont {Prance},
  \citenamefont {Mohr}, \citenamefont {Koh}, \citenamefont {Gamble},
  \citenamefont {Wu}, \citenamefont {Savage}, \citenamefont {Lagally},
  \citenamefont {Friesen}, \citenamefont {Coppersmith},\ and\ \citenamefont
  {Eriksson}}]{Shi_PRB13}%
  \BibitemOpen
  \bibfield  {author} {\bibinfo {author} {\bibfnamefont {Z.}~\bibnamefont
  {Shi}}, \bibinfo {author} {\bibfnamefont {C.~B.}\ \bibnamefont {Simmons}},
  \bibinfo {author} {\bibfnamefont {D.~R.}\ \bibnamefont {Ward}}, \bibinfo
  {author} {\bibfnamefont {J.~R.}\ \bibnamefont {Prance}}, \bibinfo {author}
  {\bibfnamefont {R.~T.}\ \bibnamefont {Mohr}}, \bibinfo {author}
  {\bibfnamefont {T.~S.}\ \bibnamefont {Koh}}, \bibinfo {author} {\bibfnamefont
  {J.~K.}\ \bibnamefont {Gamble}}, \bibinfo {author} {\bibfnamefont
  {X.}~\bibnamefont {Wu}}, \bibinfo {author} {\bibfnamefont {D.~E.}\
  \bibnamefont {Savage}}, \bibinfo {author} {\bibfnamefont {M.~G.}\
  \bibnamefont {Lagally}}, \bibinfo {author} {\bibfnamefont {M.}~\bibnamefont
  {Friesen}}, \bibinfo {author} {\bibfnamefont {S.~N.}\ \bibnamefont
  {Coppersmith}}, \ and\ \bibinfo {author} {\bibfnamefont {M.~A.}\ \bibnamefont
  {Eriksson}},\ }\href {\doibase 10.1103/PhysRevB.88.075416} {\bibfield
  {journal} {\bibinfo  {journal} {Phys. Rev. B}\ }\textbf {\bibinfo {volume}
  {88}},\ \bibinfo {pages} {075416} (\bibinfo {year} {2013})}\BibitemShut
  {NoStop}%
\bibitem [{\citenamefont {Reinisch}\ and\ \citenamefont
  {Heuer}(2006)}]{Reinisch06}%
  \BibitemOpen
  \bibfield  {author} {\bibinfo {author} {\bibfnamefont {J.}~\bibnamefont
  {Reinisch}}\ and\ \bibinfo {author} {\bibfnamefont {A.}~\bibnamefont
  {Heuer}},\ }\href {https://arxiv.org/abs/cond-mat/0609693} {\bibfield
  {journal} {\bibinfo  {journal} {arXiv:cond-mat/0609693}\ } (\bibinfo {year}
  {2006})}\BibitemShut {NoStop}%
\bibitem [{\citenamefont {Biswas}\ and\ \citenamefont
  {Li}(1999)}]{biswasHydrogenFlipModel1999}%
  \BibitemOpen
  \bibfield  {author} {\bibinfo {author} {\bibfnamefont {R.}~\bibnamefont
  {Biswas}}\ and\ \bibinfo {author} {\bibfnamefont {Y.-P.}\ \bibnamefont
  {Li}},\ }\href {\doibase 10.1103/PhysRevLett.82.2512} {\bibfield  {journal}
  {\bibinfo  {journal} {Physical Review Letters}\ }\textbf {\bibinfo {volume}
  {82}},\ \bibinfo {pages} {2512} (\bibinfo {year} {1999})}\BibitemShut
  {NoStop}%
\bibitem [{\citenamefont {Zwerver}\ \emph {et~al.}(2022)\citenamefont
  {Zwerver}, \citenamefont {Kr{\"a}henmann}, \citenamefont {Watson},
  \citenamefont {Lampert}, \citenamefont {George}, \citenamefont
  {Pillarisetty}, \citenamefont {Bojarski}, \citenamefont {Amin}, \citenamefont
  {Amitonov}, \citenamefont {Boter}, \citenamefont {Caudillo}, \citenamefont
  {Correas-Serrano}, \citenamefont {Dehollain}, \citenamefont {Droulers},
  \citenamefont {Henry}, \citenamefont {Kotlyar}, \citenamefont {Lodari},
  \citenamefont {L{\"u}thi}, \citenamefont {Michalak}, \citenamefont {Mueller},
  \citenamefont {Neyens}, \citenamefont {Roberts}, \citenamefont {Samkharadze},
  \citenamefont {Zheng}, \citenamefont {Zietz}, \citenamefont {Scappucci},
  \citenamefont {Veldhorst}, \citenamefont {Vandersypen},\ and\ \citenamefont
  {Clarke}}]{Zwerver2022}%
  \BibitemOpen
  \bibfield  {author} {\bibinfo {author} {\bibfnamefont {A.~M.~J.}\
  \bibnamefont {Zwerver}}, \bibinfo {author} {\bibfnamefont {T.}~\bibnamefont
  {Kr{\"a}henmann}}, \bibinfo {author} {\bibfnamefont {T.~F.}\ \bibnamefont
  {Watson}}, \bibinfo {author} {\bibfnamefont {L.}~\bibnamefont {Lampert}},
  \bibinfo {author} {\bibfnamefont {H.~C.}\ \bibnamefont {George}}, \bibinfo
  {author} {\bibfnamefont {R.}~\bibnamefont {Pillarisetty}}, \bibinfo {author}
  {\bibfnamefont {S.~A.}\ \bibnamefont {Bojarski}}, \bibinfo {author}
  {\bibfnamefont {P.}~\bibnamefont {Amin}}, \bibinfo {author} {\bibfnamefont
  {S.~V.}\ \bibnamefont {Amitonov}}, \bibinfo {author} {\bibfnamefont {J.~M.}\
  \bibnamefont {Boter}}, \bibinfo {author} {\bibfnamefont {R.}~\bibnamefont
  {Caudillo}}, \bibinfo {author} {\bibfnamefont {D.}~\bibnamefont
  {Correas-Serrano}}, \bibinfo {author} {\bibfnamefont {J.~P.}\ \bibnamefont
  {Dehollain}}, \bibinfo {author} {\bibfnamefont {G.}~\bibnamefont {Droulers}},
  \bibinfo {author} {\bibfnamefont {E.~M.}\ \bibnamefont {Henry}}, \bibinfo
  {author} {\bibfnamefont {R.}~\bibnamefont {Kotlyar}}, \bibinfo {author}
  {\bibfnamefont {M.}~\bibnamefont {Lodari}}, \bibinfo {author} {\bibfnamefont
  {F.}~\bibnamefont {L{\"u}thi}}, \bibinfo {author} {\bibfnamefont {D.~J.}\
  \bibnamefont {Michalak}}, \bibinfo {author} {\bibfnamefont {B.~K.}\
  \bibnamefont {Mueller}}, \bibinfo {author} {\bibfnamefont {S.}~\bibnamefont
  {Neyens}}, \bibinfo {author} {\bibfnamefont {J.}~\bibnamefont {Roberts}},
  \bibinfo {author} {\bibfnamefont {N.}~\bibnamefont {Samkharadze}}, \bibinfo
  {author} {\bibfnamefont {G.}~\bibnamefont {Zheng}}, \bibinfo {author}
  {\bibfnamefont {O.~K.}\ \bibnamefont {Zietz}}, \bibinfo {author}
  {\bibfnamefont {G.}~\bibnamefont {Scappucci}}, \bibinfo {author}
  {\bibfnamefont {M.}~\bibnamefont {Veldhorst}}, \bibinfo {author}
  {\bibfnamefont {L.~M.~K.}\ \bibnamefont {Vandersypen}}, \ and\ \bibinfo
  {author} {\bibfnamefont {J.~S.}\ \bibnamefont {Clarke}},\ }\href {\doibase
  10.1038/s41928-022-00727-9} {\bibfield  {journal} {\bibinfo  {journal}
  {Nature Electronics}\ }\textbf {\bibinfo {volume} {5}},\ \bibinfo {pages}
  {184} (\bibinfo {year} {2022})}\BibitemShut {NoStop}%
\bibitem [{\citenamefont {Kafanov}\ \emph {et~al.}(2008)\citenamefont
  {Kafanov}, \citenamefont {Brenning}, \citenamefont {Duty},\ and\
  \citenamefont {Delsing}}]{kafanovChargeNoiseSingleelectron2008}%
  \BibitemOpen
  \bibfield  {author} {\bibinfo {author} {\bibfnamefont {S.}~\bibnamefont
  {Kafanov}}, \bibinfo {author} {\bibfnamefont {H.}~\bibnamefont {Brenning}},
  \bibinfo {author} {\bibfnamefont {T.}~\bibnamefont {Duty}}, \ and\ \bibinfo
  {author} {\bibfnamefont {P.}~\bibnamefont {Delsing}},\ }\href {\doibase
  10.1103/PhysRevB.78.125411} {\bibfield  {journal} {\bibinfo  {journal}
  {Physical Review B}\ }\textbf {\bibinfo {volume} {78}},\ \bibinfo {pages}
  {125411} (\bibinfo {year} {2008})}\BibitemShut {NoStop}%
\bibitem [{\citenamefont {Rojas-Arias}\ \emph {et~al.}(2023)\citenamefont
  {Rojas-Arias}, \citenamefont {Noiri}, \citenamefont {Stano}, \citenamefont
  {Nakajima}, \citenamefont {Yoneda}, \citenamefont {Takeda}, \citenamefont
  {Kobayashi}, \citenamefont {Sammak}, \citenamefont {Scappucci}, \citenamefont
  {Loss},\ and\ \citenamefont {Tarucha}}]{Rojas23}%
  \BibitemOpen
  \bibfield  {author} {\bibinfo {author} {\bibfnamefont {J.~S.}\ \bibnamefont
  {Rojas-Arias}}, \bibinfo {author} {\bibfnamefont {A.}~\bibnamefont {Noiri}},
  \bibinfo {author} {\bibfnamefont {P.}~\bibnamefont {Stano}}, \bibinfo
  {author} {\bibfnamefont {T.}~\bibnamefont {Nakajima}}, \bibinfo {author}
  {\bibfnamefont {J.}~\bibnamefont {Yoneda}}, \bibinfo {author} {\bibfnamefont
  {K.}~\bibnamefont {Takeda}}, \bibinfo {author} {\bibfnamefont
  {T.}~\bibnamefont {Kobayashi}}, \bibinfo {author} {\bibfnamefont
  {A.}~\bibnamefont {Sammak}}, \bibinfo {author} {\bibfnamefont
  {G.}~\bibnamefont {Scappucci}}, \bibinfo {author} {\bibfnamefont
  {D.}~\bibnamefont {Loss}}, \ and\ \bibinfo {author} {\bibfnamefont
  {S.}~\bibnamefont {Tarucha}},\ }\href {http://arxiv.org/abs/2302.11717}
  {\bibfield  {journal} {\bibinfo  {journal} {arXiv:2302.11717}\ } (\bibinfo
  {year} {2023})}\BibitemShut {NoStop}%
\bibitem [{\citenamefont {K\k{e}pa}\ \emph {et~al.}(2024)\citenamefont
  {K\k{e}pa}, \citenamefont {Focke}, \citenamefont {Cywi\'nski},\ and\
  \citenamefont {Krzywda}}]{repo}%
  \BibitemOpen
  \bibfield  {author} {\bibinfo {author} {\bibfnamefont {M.}~\bibnamefont
  {K\k{e}pa}}, \bibinfo {author} {\bibfnamefont {N.}~\bibnamefont {Focke}},
  \bibinfo {author} {\bibfnamefont {{\L}.}~\bibnamefont {Cywi\'nski}}, \ and\
  \bibinfo {author} {\bibfnamefont {J.~A.}\ \bibnamefont {Krzywda}},\ }\href
  {https://doi.org/10.24435/materialscloud:mx-0w} {\bibfield  {journal}
  {\bibinfo  {journal} {Materials Cloud Archive}\ }\textbf {\bibinfo {volume}
  {2024.118}} (\bibinfo {year} {2024})}\BibitemShut {NoStop}%
\end{thebibliography}%
\end{document}